\documentclass{ws-rv9x6-mod}

\usepackage{subfigure}
\usepackage{hyperref}

\usepackage{ws-rv-van}

\newcommand{\half}{{\tfrac{1}{2}}}

\newcommand{\tvec}[1]{\boldsymbol{#1}}

\newcommand{\ms}{\mskip 1.5mu}
\newcommand{\bs}{\mskip -1.5mu}

\newcommand{\prn}[2]{{}^{#1} #2}       
\newcommand{\prb}[2]{{}^{#1}\bs #2}    
\newcommand{\prl}[2]{{}^{#1}\! #2}     

\begin{document}
\chapter[Double parton scattering theory overview]{Double Parton Scattering --
  Theory Overview}

\author[M. Diehl and J. R. Gaunt]{Markus Diehl$^{*}$ and Jonathan R.
  Gaunt$^{\dagger}$} 

\address{$^{*}$Deutsches Elektronen-Synchroton DESY, \\
  Notkestra{\ss}e 85, 22607 Hamburg, Germany}

\address{$^{\dagger}$CERN Theory Division, \\
  1211 Geneva 23, Switzerland}

\begin{abstract}
The dynamics of double hard scattering in proton-proton collisions is
quite involved compared with the familiar case of single hard scattering.
In this contribution, we review our theoretical understanding of double
hard scattering and of its interplay with other reaction mechanisms.
\end{abstract}

\body

\section{Introduction}

The most familiar mechanism for hard processes in proton-proton collisions
is single parton scattering (SPS): two partons, one from each proton,
undergo a hard scattering that produces heavy particles or particles with
high transverse momenta.  For the cross section one then has a
factorisation formula, containing a parton distribution function (PDF) for
each proton and a parton-level cross section for the hard subprocess.
Double parton scattering (DPS) occurs if in the same proton-proton
collision two partons in each proton initiate two separate hard scattering
processes.  The corresponding factorisation formula contains two
parton-level cross sections and a double parton distribution (DPD) for
each proton.  The two hard scatters are separated by a finite distance
$\tvec{y}$ in the plane transverse to the colliding proton momenta, so
that a DPD depends not only on the momentum fractions $x_1$ and $x_2$ of
two partons, but also on the transverse distance $\tvec{y}$ between them.
Very roughly, DPDs should grow like the square of two ordinary PDFs when
$x_1$ and $x_2$ become small.  The importance of DPS compared with SPS is
hence increased in this small $x$ region, which for a given final state
becomes more and more important with growing collision energy.

The single and double parton distributions just described are integrated
over transverse parton momenta; they are often called ``collinear''
distributions, and the associated formalism is called ``collinear
factorisation''.  The information on transverse parton momenta is retained
in so-called TMDs (transverse momentum dependent distributions).  The
corresponding TMD factorisation formulae allow one to compute cross
sections differential in the transverse momentum $\tvec{q}$ of a heavy
particle (e.g.\ a $Z$ or a Higgs boson) in the region where $\tvec{q}$ is
much smaller than the boson mass.  The TMD concept can be extended to DPS
processes, for instance to describe the region of low transverse boson
momenta $\tvec{q}_1$ and $\tvec{q}_2$ in $W^+ W^+$ or $H Z$ production.
This is especially valuable because the importance of DPS compared with
SPS is much higher in the cross section for measured small $\tvec{q}_1$
and $\tvec{q}_2$ than it is in the integrated cross section.

Factorisation for SPS processes has been derived within QCD to a high
level of rigour, as reviewed for instance in
Ref.~\refcite{Collins:2011zzd}.  It is an ongoing effort to bring
factorisation for DPS to a comparable standard.  In the present
contribution, we review the status of this effort.  Note that we are
discussing so-called ``hard scattering factorisation'' here, which is
based on separating dynamics at different distance scales.  This is
distinct from ``high-energy'' or ``small $x$ factorisation'', where the
separation criterion is rapidity.  Some discussion of this concept
in the context of DPS is given in Ref.~\refcite{Lonnblad:book}.


\section{Cross section formula}
\label{sec:cross-sect}

Let us start with a main theory result: the cross section formula for DPS.
Consider the production of two particles with invariant masses $Q_1$,
$Q_2$ and transverse momenta $\tvec{q}_1$, $\tvec{q}_2$.  We require that
$Q_1$ and $Q_2$ be large and generically denote their size by $Q$.
Instead of a heavy particle, one may also have a system of particles with
large invariant mass, for instance a dijet.

Collinear factorisation allows us to compute the cross section integrated
over $\tvec{q}_1$ and $\tvec{q}_2$\,:
\begin{align}
  \label{coll-Xsect-master}
& \frac{d\sigma_{\text{DPS}}}{dx_1\, dx_2\,
          d\bar{x}_1\, d\bar{x}_2}
  = \frac{1}{C}  \!\! \sum_{a_1 a_2 b_1 b_2}
    \int_{x_1}^{1-x_2} \frac{dx_1'}{x_1'} 
    \int_{x_2}^{1-x_1'} \frac{dx_2'}{x_2'}
    \int_{\bar{x}_1}^{1-\bar{x}_2} \frac{d\bar{x}_1'}{\bar{x}_1'} 
    \int_{\bar{x}_2}^{1-\bar{x}_1'} \frac{d\bar{x}_2'}{\bar{x}_2'}
\nonumber \\[0.4em]
 & \qquad \times \sum_{R}
     \prn{R}{\hat{\sigma}}_{a_1 b_1}(x_1'\ms \bar{x}_1' s, \mu_1^2)\,
     \prn{R}{\hat{\sigma}}_{a_2 b_2}(x_2'\ms \bar{x}_2'\ms s, \mu_2^2)\,
\nonumber \\
 & \qquad \times \int d^2\tvec{y}\; \Phi^2(y \nu)\;
     \prb{R}{F}_{b_1 b_2}(\bar{x}_i',\tvec{y};\mu_i^{}, \bar{\zeta}) \,
     \prb{R}{F}_{a_1 a_2}(x_i',\tvec{y};\mu_i^{}, \zeta) \,.
\end{align}
Note that we use boldface for any vector $\tvec{w}$ in the transverse
plane and denote its length by $w = |\tvec{w}|$.  There are strong
indications\cite{Rogers:2010dm} that TMD factorisation in SPS works only
for the production of colourless particles, so that we make the same
restriction for DPS.  The differential cross section for transverse
momenta $|\tvec{q}_1|, |\tvec{q}_2| \sim q_T$ much smaller than $Q$ reads
\begin{align}
  \label{TMD-Xsect-master}
& \frac{d\sigma_{\text{DPS}}}{dx_1\, dx_2\,
          d\bar{x}_1\, d\bar{x}_2\, d^2\tvec{q}_1\, d^2\tvec{q}_2}
  = \frac{1}{C}\, \sum_{a_1 a_2 b_1 b_2} \!\!\!
     \hat{\sigma}_{a_1 b_1}(Q_1^2, \mu_1^2)\,
     \hat{\sigma}_{a_2 b_2}(Q_2^2, \mu_2^2)\,
\nonumber \\
 & \qquad \times
     \int d^2\tvec{y}\; \frac{d^2\tvec{z}_1}{(2\pi)^2}\,
          \frac{d^2\tvec{z}_2}{(2\pi)^2}\;
       e^{-i (\tvec{q}_1^{} \tvec{z}_1^{} +\tvec{q}_2^{} \tvec{z}_2^{})}\,
              \Phi(y_+ \nu)\ms \Phi(y_- \nu)
\nonumber \\[0.4em]
 & \qquad \times \sum_{R}
     \prb{R}{F}_{b_1 b_2}(\bar{x}_i,\tvec{z}_i,\tvec{y};\mu_i, \bar{\zeta}) \,
     \prb{R}{F}_{a_1 a_2}(x_i,\tvec{z}_i,\tvec{y};\mu_i, \zeta) \,.
\end{align}
These formulae are quite complex.  In the following we briefly explain
their different ingredients, and the physics behind them.

We begin with the simplest ones.  The variables $x_i$ and $\bar{x}_i$ are
given by
\begin{align}
  \label{mom-fracs}
x_i &= Q_i\, e^{Y_i}/\sqrt{s} \,,
&
\bar{x}_i &= Q_i\, e^{-Y_i}/\sqrt{s} \,,
&
(i &= 1,2)
\end{align}
where $Y_i$ is the centre-of-mass rapidity of the system $i$ and
$\sqrt{s}$ the overall collision energy.  $C$ is a combinatorial factor,
equal to $2$ if the systems $1$ and $2$ are identical, and equal to $1$
otherwise.

The parton-level cross sections $\hat{\sigma}$ are precisely the same as
the ones in the corresponding SPS cross sections, except for the
superscript $R$ in \eqref{coll-Xsect-master}, which will be explained
below.  They include the effects of hard QCD radiation in the process.
In TMD factorisation, $\hat{\sigma}$ receives only virtual corrections,
since hard real radiation tends to knock $\tvec{q}_i$ out of the region
$q_T \ll Q_i$.  As a consequence, the momentum fractions of the partons
entering the hard subprocesses are fixed to $x_i$ and $\bar{x}_i$ by
external kinematics.  In collinear factorisation, $\hat{\sigma}$ includes
real emission, which allows for momentum fractions $x'{}_{\!\! i} \ge x_i$
and $\bar{x}'{}_{\!\! i} \ge \bar{x}_i$.

The joint distribution of two partons in a proton is quantified by the
DPDs $F$, which have two labels $a_i$ for the parton type, two momentum
fraction arguments $x_i$, and two factorisation scales $\mu_i$ (they can
be chosen separately, which is useful if $Q_1$ and $Q_2$ are of different
size).  In the TMD case there are two transverse position arguments
$\tvec{z}_i$, which are Fourier conjugate to the transverse parton momenta
$\tvec{k}_i$.  The structure $\int d^2\tvec{z}_i\; e^{-i \tvec{q}{}_i
  \tvec{z}_i}\, F(\bar{x}_i, \tvec{z}_i, \cdots)\, F(x_i, \tvec{z}_i,
\cdots)$ in \eqref{TMD-Xsect-master} is the same as in the corresponding
factorisation formula for SPS -- in momentum space it corresponds to a
convolution product $\int d^2\tvec{k}_i\; F(\bar{x}_i,
\tvec{q}{}_i-\tvec{k}_i, \cdots)\, F(x_i, \tvec{k}_i, \cdots)$.

As already mentioned, a DPD also depends on the distance $\tvec{y}$, which
in collinear factorisation literally corresponds to the transverse
distance between the two active partons in the proton, and thus to the
distance between the two hard-scattering processes.  In the TMD case,
$\tvec{y}$ corresponds to the average distance between the partons in the
scattering amplitude and its conjugate, as can be seen in \eqref{tmd-dpds}
and \eqref{quark-ops} below.  Notice that in the cross section, $\tvec{y}$
is not Fourier conjugate to any observable momentum, unlike~$\tvec{z}_i$.

\begin{figure}
\begin{center}
\subfigure[]{\includegraphics[height=0.303\textwidth]{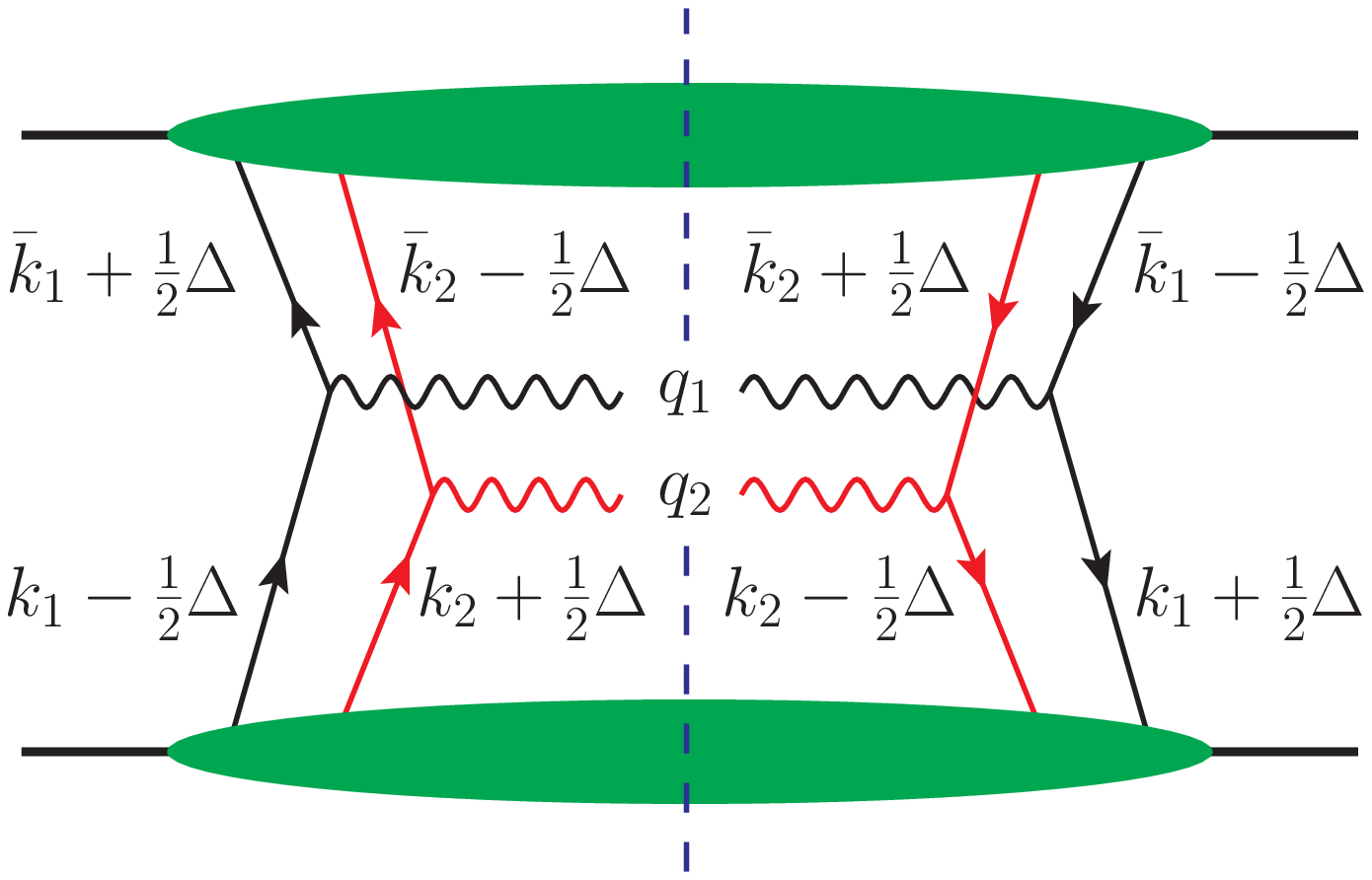}}
\subfigure[]{\includegraphics[height=0.299\textwidth]{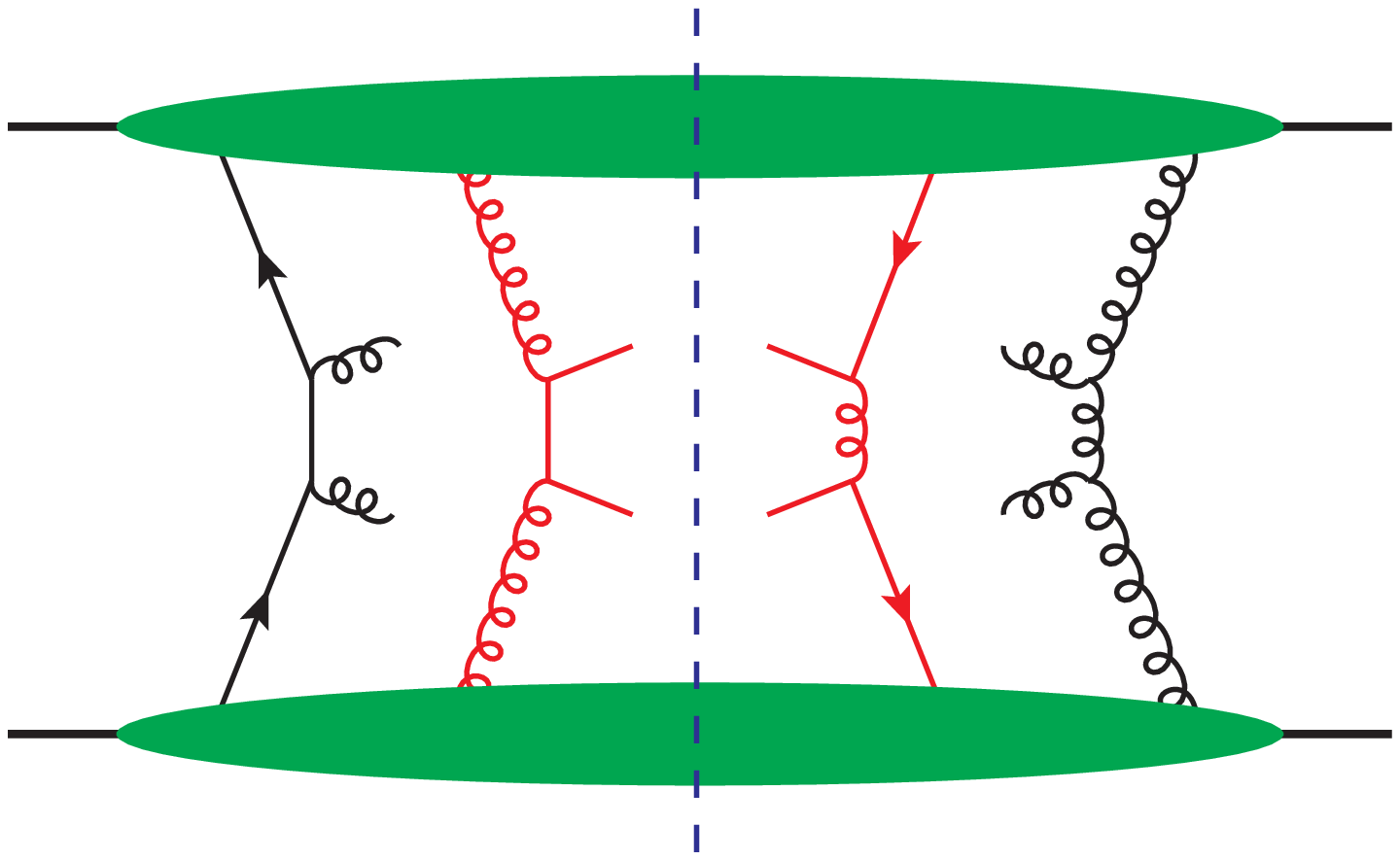}}
\caption{\label{fig:double} (a) Tree level graph for the production of two
  electroweak gauge bosons by DPS (often called double Drell-Yan).  The
  blobs represent DPDs.  The graph is for the cross section, with the
  vertical line indicating the final state cut.  (b)~Graph for double
  dijet production with DPDs for quark-gluon interference.}
\end{center}
\end{figure}

It is instructive to see how the distance $\tvec{y}$ emerges from the
analysis of Feynman graphs in momentum space.  The longitudinal momentum
fractions of each parton are fixed by the final state kinematics and thus
must be equal in the scattering amplitude and its conjugate.  By
contrast, the transverse parton momenta can differ by an amount
$\tvec{\Delta}$ or~$-\tvec{\Delta}$ as shown in \fref{fig:double}(a).  The
momentum mismatch for the first and the second parton is opposite in sign,
so that the transverse momentum of the spectator partons is the same in
the amplitude and its conjugate.  Since this momentum mismatch is
not observable, one has an integral of the form $\int d^2\tvec{\Delta}\;
F(\bar{x}_i,-\tvec{\Delta},\cdots)\, F(x_i,\tvec{\Delta},\cdots)$ in the
cross section.  A Fourier transform from $\tvec{\Delta}$ to $\tvec{y}$
gives the form shown in \eqref{coll-Xsect-master} and
\eqref{TMD-Xsect-master}.  More detail on the tree-level derivation of the
factorised structure in collinear factorisation is given in
Ref.~\refcite{Treleani:book}.

Let us now turn to the quantum numbers of the partons.  Even in an
unpolarised proton, two extracted partons can have correlations between
their polarisations.  The labels $a_i, b_i$ in the cross section formulae
refer not only to the type of the parton but also to its polarisation, and
one must sum over all allowed combinations.  An example for a
polarisation dependent DPD is $F_{\Delta q \Delta q}$, which corresponds
to the difference of distributions for two quarks with equal helicities
and for two quarks with opposite helicities.

Not only the transverse momentum of a parton can differ between the
amplitude and its conjugate, but also its colour.  The different possible
colour combinations in DPDs and the parton-level cross sections are
specified by the label $R$.  For the production of colourless particles
there is only one possible colour structure for $\hat{\sigma}$, which
hence requires no index $R$ in the TMD formula \eqref{TMD-Xsect-master}.
Let us explain the meaning of $R$ for the simplified setting of the tree
graph in \fref{fig:double}(a).  In each DPD, one can couple the two parton
lines with momentum fraction $x_1$ ($\bar{x}_1$) to be in the colour
representation $R = 1,8,\ldots$ (the two other lines then are in the
conjugate representation, because all four lines must couple to an overall
singlet).  In colour singlet distributions $\prb{1}{F}$, partons with
equal momentum fractions thus have equal colour~-- this is the only
possible combination for single parton distributions.  Colour non-singlet
DPDs describe colour correlations.  At the end of \sref{sec:sudakov} we
will see that they are suppressed by Sudakov logarithms if the scale of
the hard process is large.

Finally, there also exist DPDs describing the interference between
different parton types in the amplitude and its conjugate, be it between
different quark flavours, between quarks and antiquarks, or between quarks
and gluons.  For ease of notation, they are not included in the cross
section formulae \eqref{coll-Xsect-master} and \eqref{TMD-Xsect-master}.
An example for quark-gluon interference in double dijet production is
given in \fref{fig:double}(b).  Parton type interference distributions do
not have any dynamical cross talk with gluon DPDs, which have the
strongest enhancement at small $x_i$.  In many situations, one can
therefore expect them to play only a minor role.  A detailed discussion of
correlations in DPDs can be found in Ref.~\refcite{Kasemets:book}.

DPDs can be defined via operator matrix elements, which provides a solid
field theoretical basis for their investigation.  For a double quark TMD
one writes
\begin{align}
\label{tmd-dpds}
& \prb{R}{F}_{a_1 a_2}(x_i,\tvec{z}_i,\tvec{y}; \mu_i,\zeta) 
 = 2p^+ \int dy^-\, \frac{dz^-_1}{2\pi}\, \frac{dz^-_2}{2\pi}\,
   e^{i\ms ( x_1^{} z_1^- + x_2^{} z_2^-)\ms p^+}
\nonumber \\
& \qquad\qquad \times \langle p \ms|\, \mathcal{O}_{a_2}(0,z_2)\, 
	\mathcal{O}_{a_1}(y,z_1) \,|\ms p \rangle \times 
     \{ \text{soft factor} \} \,,
\end{align}
where we use light-cone coordinates $w^{\pm} = (w^0 \pm w^3)/\sqrt{2}$ for
any four-vector $w^\mu$.  It is understood that $\tvec{p}=\tvec{0}$ and
that the proton spin is averaged over.  The bilinear operators
$\mathcal{O}$ are the same as in the definition of a single parton TMD.
They are given by
\begin{align}
  \label{quark-ops}
\mathcal{O}_{a}(y,z)
&= \bar{q}\bigl( y - \half z \bigr)\ms W^\dagger \bigl(y-\half z \bigr) \,
	\Gamma_a \, W \bigl(y+\half z \bigr)\ms q\bigl( y + \half z \bigr)
\Big|_{z^+ = y^+_{\phantom{i}} = 0}
\end{align}
with a past-pointing light-like Wilson line
\begin{align}
  \label{WL-def}
W(\xi) = \operatorname{P} 
\exp \biggl[\, i g\ms t^a \int_{0}^{\infty}
    \!\!ds\; n A^a(\xi-s n) \biggr] \,,
\end{align}
where $\operatorname{P}$ denotes path-ordering and $n$ is a light-like
vector ($n^- = 1$, $n^+ = 0$, $\tvec{n}=\tvec{0}$).  The dynamical origin
of this Wilson line is explained in \sref{sec:sudakov}.  $\Gamma_a$ is a
Dirac matrix and determines the quark polarisation.  In particular,
unpolarised quarks correspond to $\Gamma_q = \half \gamma^+$, and
longitudinal quark polarisation is described by $\Gamma_{\Delta q} = \half
\gamma^+ \gamma_5$.

The ``soft factor'' in \eqref{tmd-dpds} originates from soft gluon
exchange in the physical scattering process and gives rise to the
dependence on a parameter $\zeta$, as explained in \sref{sec:sudakov}.
Such a dependence is already present in single parton TMDs.  Moreover, the
operator \eqref{quark-ops} and the soft factor contain ultraviolet
divergences, which require renormalisation.  This brings in the dependence
on the renormalisation scales $\mu_i$.  Finally, the dependence of the DPD
on $R$ arises from the colour indices of the operators $(\bar{q}\,
W^\dagger)_{i'}$ and $(W q)_{i}$ in \eqref{quark-ops} and from the soft
factor.  Again, more detail is given in \sref{sec:sudakov}.

The preceding discussion can be repeated for antiquarks or gluons, with
different operators $\mathcal{O}_a$.
The definition of collinear DPDs $F_{a_1 a_2}(x_i,\tvec{y};\mu_i, \zeta)$
reads as in \eqref{tmd-dpds} but with $\tvec{z}_i = \tvec{0}$.  Note that
in the colour non-singlet case, the soft factor and the dependence on
$\zeta$ do not drop out in the collinear case.  Putting $\tvec{z}_i$ to
zero introduces additional ultraviolet divergences, so that the
renormalisation and hence the $\mu_i$ dependence is quite different
between TMDs and collinear distributions, as we will see later.

The role of the function $\Phi$ in \eqref{coll-Xsect-master} and
\eqref{TMD-Xsect-master} will be explained in \sref{sec:split}.  It is
closely related to the fact that the cross section of a physical process
receives not only contributions from DPS, but also from SPS and possibly
other mechanisms.  In the next section, we give an overview of these.


\section{Power behaviour}
\label{sec:power}

The factorisation of cross sections into perturbative hard-scattering
subprocesses and nonperturbative quantities like parton distributions is
based on an expansion in the small parameter $\Lambda/Q$.  Here $Q$
denotes the scale of the hard scattering and $\Lambda$ a typical hadronic
scale.  For simplicity, we treat the size of the transverse momenta
$\tvec{q}_1$ and $\tvec{q}_2$ in TMD factorisation as order $\Lambda$
here.  The case where they are much larger than a hadronic scale (but
still much smaller than $Q$) is discussed in \sref{sec:high-qt}.

Dimensional analysis of the TMD factorisation formulae for SPS and DPS
reveals that the two mechanisms have the same power behaviour:
\begin{align}
  \label{pow-TMD}
\frac{d\sigma_{\text{SPS}}}{d^2 \tvec{q}_1\, d^2\tvec{q}_2}
  & \sim \frac{d\sigma_{\text{DPS}}}{d^2 \tvec{q}_1\, d^2\tvec{q}_2}
    \sim \frac{1}{\Lambda^2 Q^4} \,.
\end{align}
The situation changes if one integrates over $\tvec{q}_1$ and
$\tvec{q}_2$.  In DPS both are of order $\Lambda$ since they originate
from the transverse momenta of partons inside the colliding protons.  In
SPS this holds only for the sum $\tvec{q}_1 + \tvec{q}_2$, whilst the
individual momenta $\tvec{q}_1$ and $\tvec{q}_2$ (and thus their
difference) are only limited by the available phase space and can hence be
of order $Q$.  One thus obtains for the integrated cross sections
\begin{align}
  \label{pow-coll}
  \sigma_{\text{SPS}} & \sim 1/Q^2 \,,
&
  \sigma_{\text{DPS}} & \sim \Lambda^2 /Q^4 \,,
\end{align}
where DPS has become power suppressed because it populates a smaller phase
space.  However, DPS can still be important in this case, for instance if
SPS is suppressed by coupling constants (the production of $W^+ W^+$ or
$W^- W^-$ is a prominent example).  Generically, DPS is enhanced if the
momentum fractions $x$ in the hard scattering subprocesses become small,
as already noted in the introduction.

\begin{figure}
\begin{center}
\subfigure[]{\includegraphics[width=0.31\textwidth]{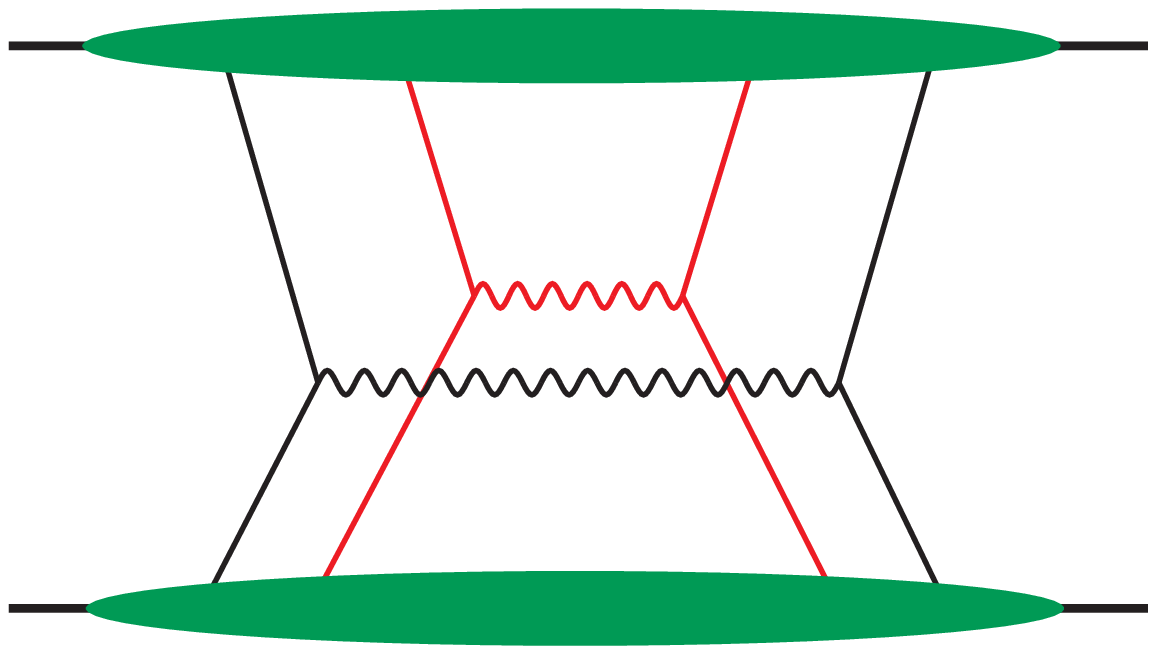}}
\hspace{0.1em}
\subfigure[]{\includegraphics[width=0.31\textwidth]{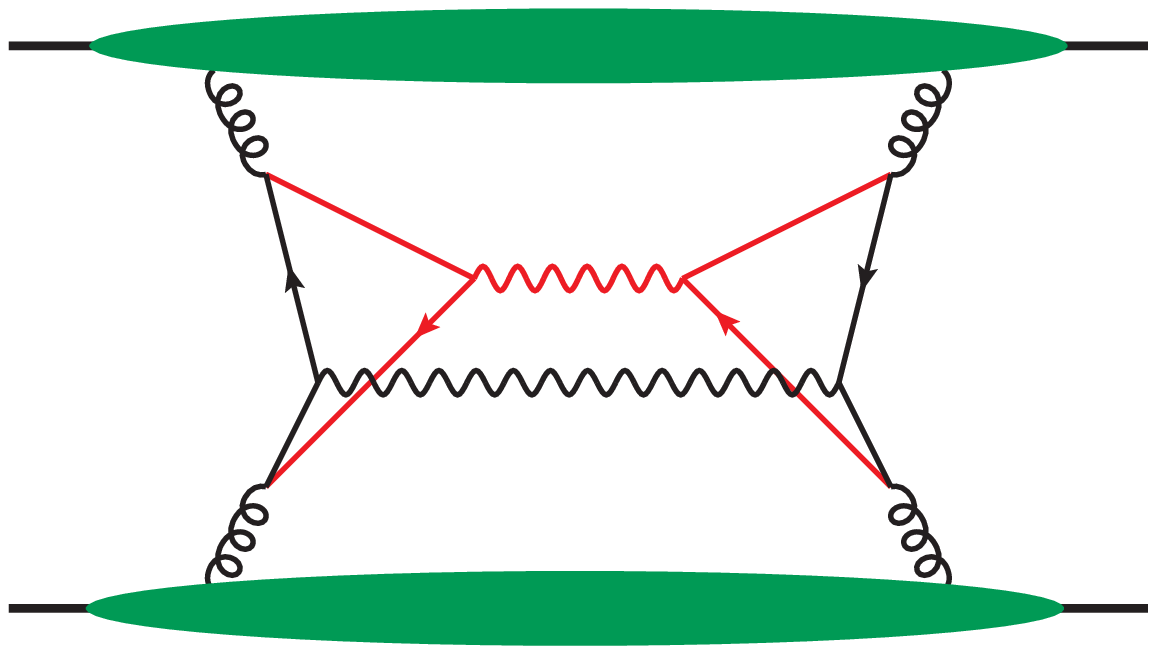}}
\hspace{0.1em}
\subfigure[]{\includegraphics[width=0.31\textwidth]{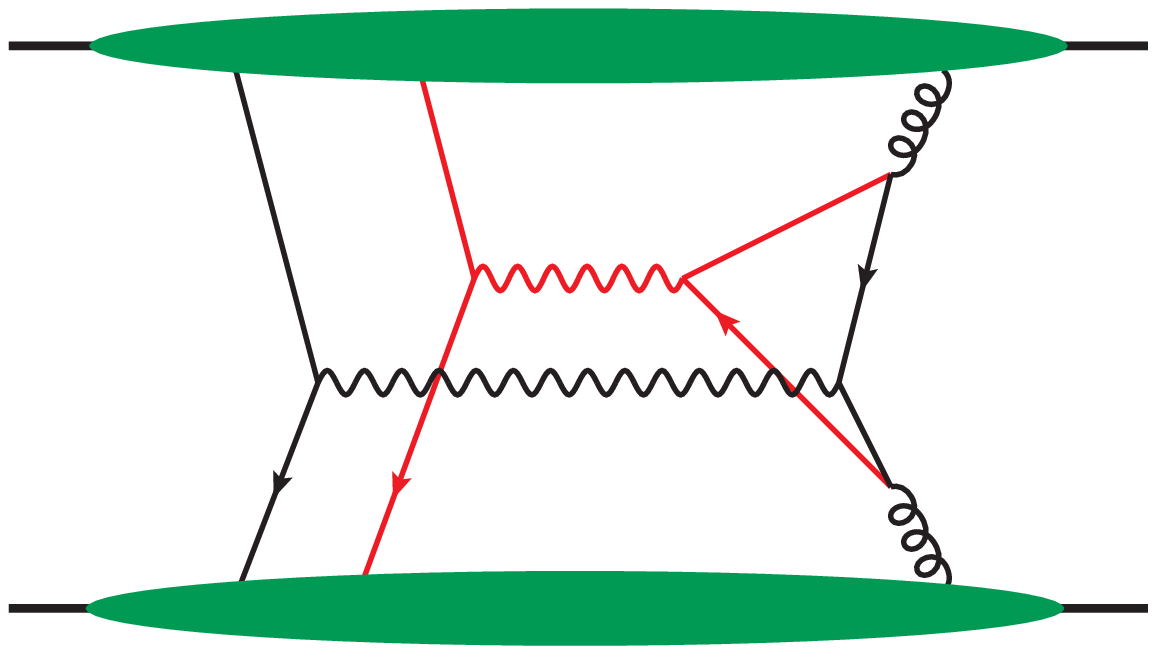}}
\\
\subfigure[]{\includegraphics[width=0.31\textwidth]{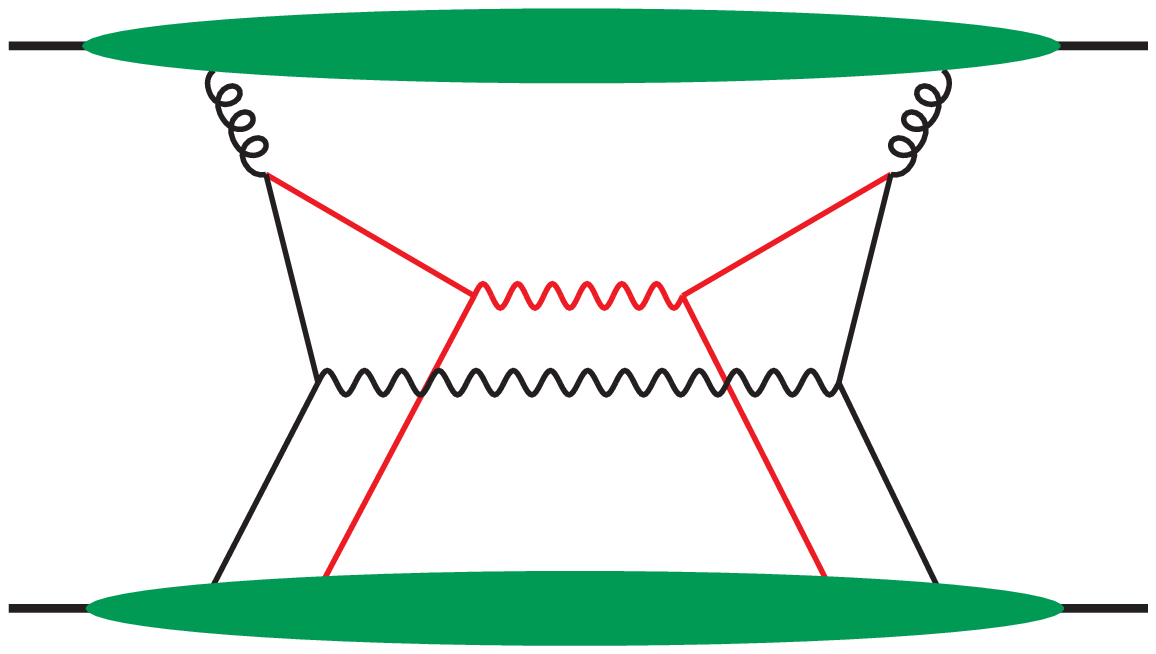}}
\hspace{0.1em}
\subfigure[]{\includegraphics[width=0.31\textwidth]{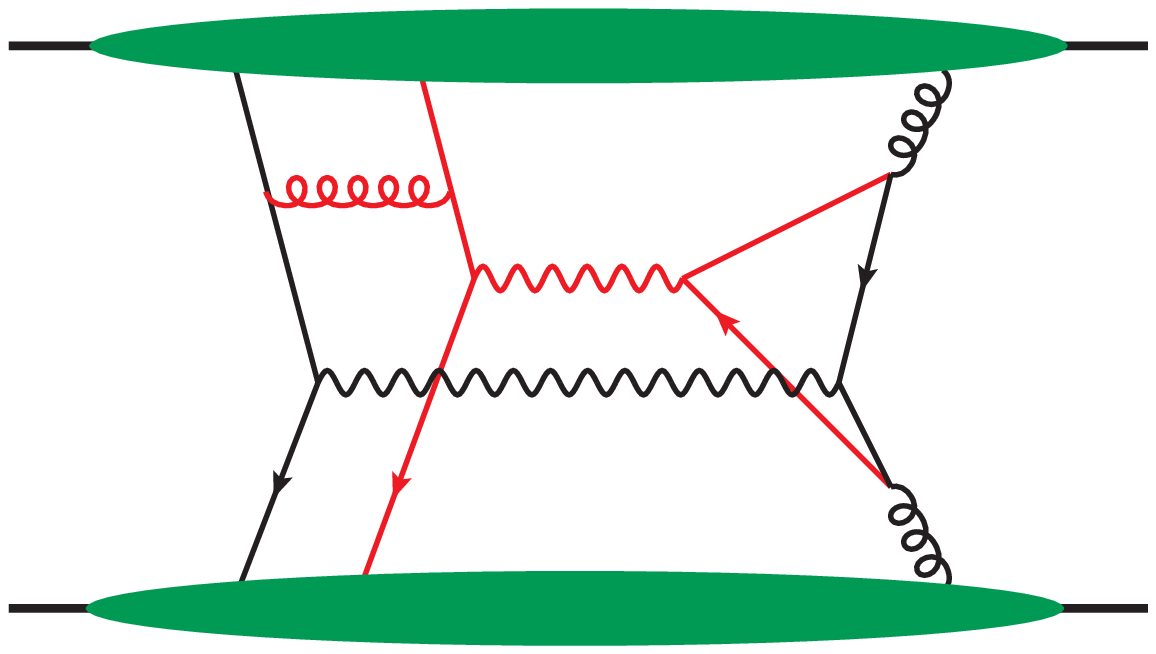}}
\caption{\label{fig:power} Different contributions to the production of
  two electroweak gauge bosons: DPS~(a), SPS (b) and their interference
  (c).  Graphs (d) and (e) involve higher-twist distributions.  Internal
  lines in the hard scattering are off shell by order $Q$.  A vertical
  line for the final state cut is not shown for simplicity.}
\end{center}
\end{figure}

There are further mechanisms that contribute at the same power to the
cross section as the terms in \eqref{pow-TMD} or \eqref{pow-coll}, as
shown in Ref.~\refcite{Diehl:2011yj}.  In TMD factorisation, the leading
power contributions are from SPS, from DPS and from the interference
between the two mechanisms.  Example graphs are given in
\fref{fig:power}(a), (b) and (c).

For the cross section integrated over $\tvec{q}_1$ and $\tvec{q}_2$, the only
leading-power contributions comes from SPS.  Suppressed by $\Lambda^2 /Q^2$
are two types of graphs in addition to DPS:
\begin{itemize}
\item graphs with a collinear twist-two distribution (i.e.\ a parton density)
  for one proton and a collinear twist-four distribution for the other one, as
  in \fref{fig:power}(d).  We refer to this as the twist-four mechanism in the
  following.
\item graphs with a collinear twist-three distribution for each proton, as
  in \fref{fig:power}(e).  This will be referred to as the twist-three
  mechanism.
\end{itemize}
As was already noted in Ref.~\refcite{Calucci:2009ea} (see also
Ref.~\refcite{Treleani:book}), the integration over $\tvec{q}_1$ and
$\tvec{q}_2$ forces all hard interactions to occur at the same transverse
position in the SPS/DPS interference, which thus becomes a special case of
the twist-three mechanism.  By contrast, in TMD factorisation the graphs
in \fref{fig:power}(d) and (e) are suppressed by $\Lambda^2/Q^2$ compared
with the SPS/DPS interference in \fref{fig:power}(c).

In an unpolarised proton, the number of possible collinear twist-three
distributions is severely restricted by helicity conservation, and only
distributions with a quark and an antiquark of opposite helicity are
allowed~\cite{Diehl:2017kgu}.  Such distributions do not have any cross
talk with gluon distributions.  One can hence expect them to lack the
small $x$ enhancement of quark or gluon DPDs, so that there is some
justification for neglecting them (in the same spirit as neglecting the
parton type interference distributions mentioned in
\sref{sec:cross-sect}).
Notice that in TMD factorisation, the twist-three distributions occurring
in the SPS/DPS interference are not subject to restrictions from parton
helicity conservation: since all three parton fields are at different
transverse positions, orbital angular momentum can compensate a mismatch
of parton helicities in this case.

\begin{figure}
\begin{center}
\includegraphics[width=0.4\textwidth]{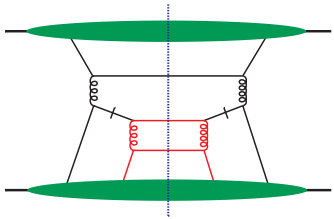}
\caption{\label{fig:resc} A graph for three-jet production by the
  twist-four mechanism that has been associated with ``rescattering'' in
  kinematics where the lines marked by a bar are on shell.  All three
  partons in the final state are understood to have large transverse
  momenta.}
\end{center}
\end{figure}

A special class of graphs for the twist-four mechanism, shown in
\fref{fig:resc}, has been associated with ``rescattering'' in
Ref.~\refcite{Paver:1984ux} (see also Ref.~\refcite{Sjostrand:book}).  Each
propagator marked by a bar in the figure has a denominator of the form $a
x - b + i\epsilon$, where $x$ is a loop variable and $a,b$ are fixed by
external kinematics.  Keeping the pole parts of each propagators and
neglecting the principal value part of the integration puts the two lines
on shell, and the process looks like one $2\to 2$ scattering followed by a
second one.  The calculation of this two-pole part in terms of two
unpolarised $2\to 2$ partonic cross sections is indeed correct if in the
twist-four distribution the quantum numbers are coupled such that partons
with equal momentum fractions are unpolarised and form a colour singlet.
However, it is not obvious that the pole parts of the loop integrations
should dominate over the principal value contributions in general
kinematics.  This may happen for jets with very large rapidities
\cite{Treleani:1994at}.  We also emphasise that
the partons marked by bars do \emph{not} physically propagate over
distances much larger than $1/Q$.  Technically speaking, their propagator
poles can be avoided by a complex contour deformation in the loop
integrals, and physically speaking one finds that the ``rescattering'' of
\fref{fig:resc} does not correspond to a classically allowed scattering
process.\cite{Diehl:2011yj}.  It is therefore inappropriate to associate
final- or initial-state parton showers to these partons.


\section{Short-distance splitting and double counting}
\label{sec:split}

\begin{figure}
\begin{center}
\subfigure[]{\includegraphics[width=0.38\textwidth]{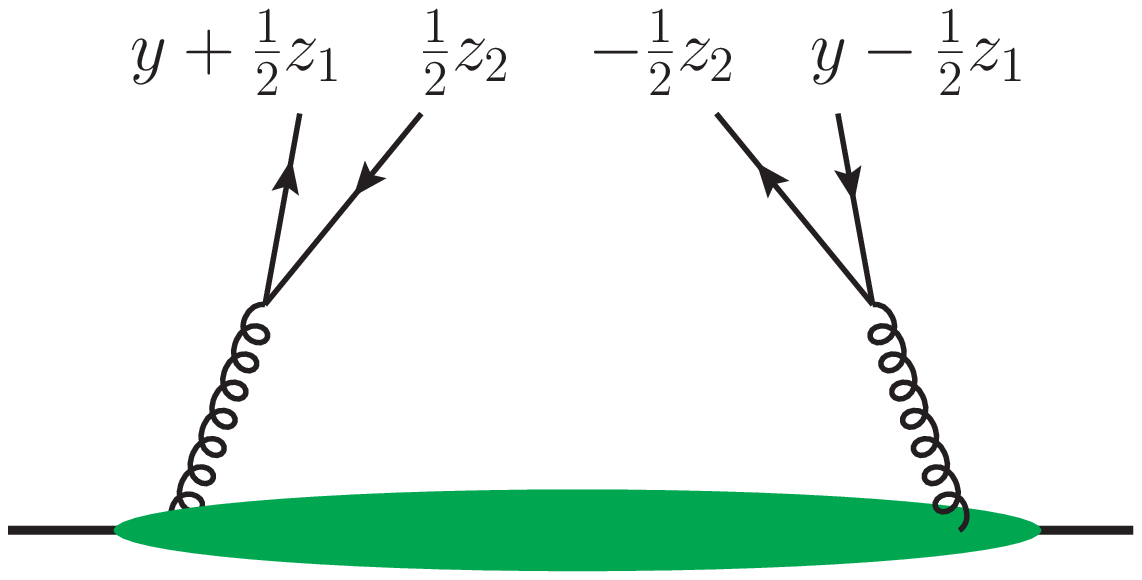}}
\subfigure[]{\includegraphics[width=0.30\textwidth]{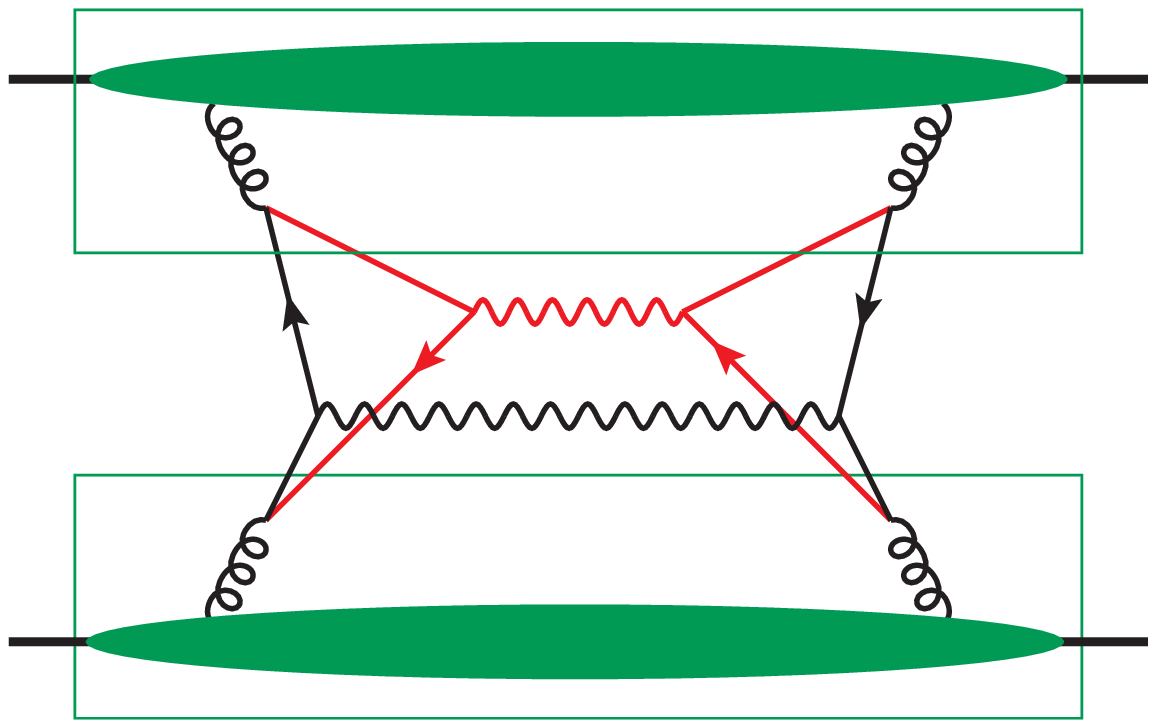}}
\subfigure[]{\includegraphics[width=0.30\textwidth]{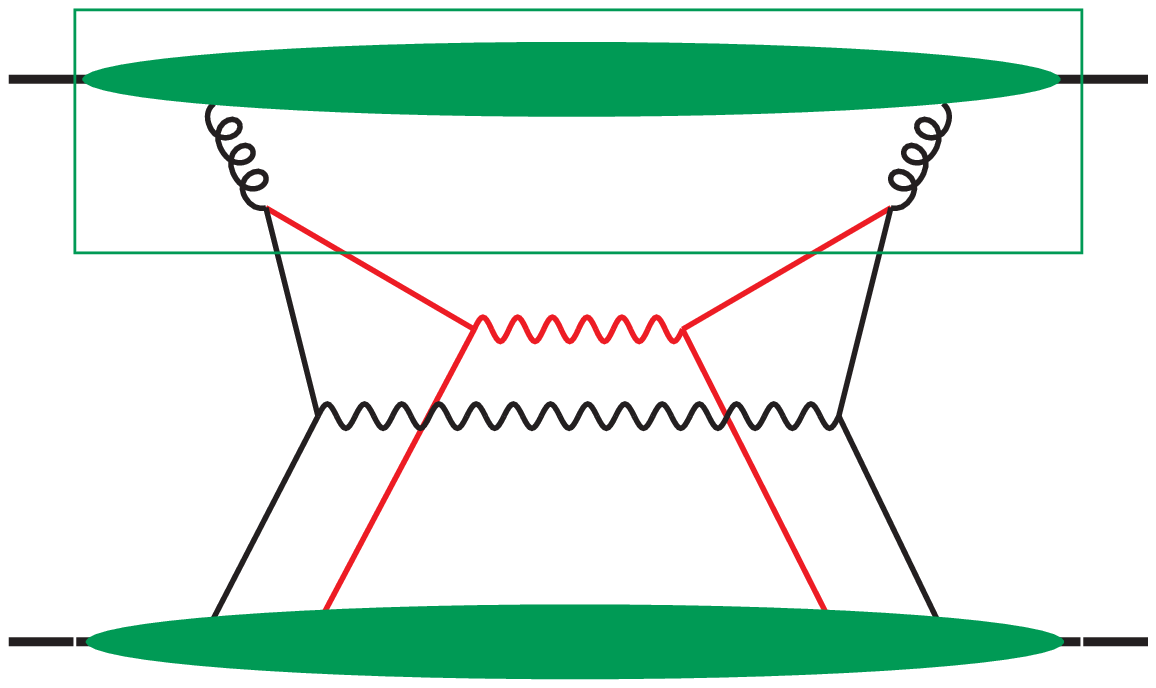}}
\caption{\label{fig:split} Splitting provides a short-distance
  contribution to DPDs (a) and gives rise to 1v1 (b) and 2v1 (c) graphs
  for the DPS cross section.  The boxes represent double parton
  distributions; partons emerging from them have virtualities much smaller
  than~$Q$.}
\end{center}
\end{figure}

At small inter-parton distances, the dominant contribution to a DPD comes
from perturbative splitting of one parton into two, as depicted in
\fref{fig:split}(a).  Let us for now concentrate on collinear DPDs.  At
leading order in $\alpha_s$, the contribution of the $1\to 2$ splitting
mechanism is easily computed and reads
\begin{align}
  \label{split-dpd}
& \prb{R}{F}_{a_1 a_2}(x_1,x_2, \tvec{y}) \big|_{\text{spl,pt}}
\nonumber \\
  &\quad = \frac{1}{{y}^2}\, \frac{\alpha_s}{2\pi^2}\,
         \prb{R}{P}_{a_0\to a_1 a_2}\biggl( \frac{x_1}{x_1+x_2} \biggr)\,
  \frac{f_{a_0}(x_1+x_2)}{x_1+x_2} \,,
\end{align}
where $f_{a_0}$ is an unpolarised PDF and $P_{a_0\to a_1 a_2}$ a splitting
function.  The $1/{y}^2$ behaviour can be deduced already by dimensional
counting.  Note that this mechanism gives strong colour and spin
correlations: chirality conservation for massless quarks results for
instance in complete anti-alignment of the quark and antiquark helicities
in $g\to q \bar{q}$.

DPDs also contain an ``intrinsic'' short-distance part, where the two partons
may be thought of as part of the ``intrinsic'' wave function of the proton.
This part is related to a twist-four distribution and only diverges
logarithmically at small $y$.  Finally, one may have a $1\to 2$ splitting only
in the amplitude or only in its conjugate: this contribution involves a
collinear twist-three distribution and lacks small $x$ enhancement, as
discussed in the previous section.

Inserting the short-distance form \eqref{split-dpd} into the cross section
formula \eqref{coll-Xsect-master}, we see that without the function $\Phi$
the integral over $\tvec{y}$ would be power divergent.  This power
divergence is associated with so-called 1v1 (1 versus 1 parton) diagrams,
in which there are $1\to 2$ splittings in both protons as shown in
\fref{fig:split}(b).  Note that this is the same graph as in
\fref{fig:power}(b), which represents a loop correction in the
leading-power SPS mechanism.  The difference is that in
\fref{fig:split}(b) the quark virtualities are understood to be much
smaller than $Q$, whereas in \fref{fig:power}(b) they are of order $Q$.
The small~$y$ divergence in the DPS formula without $\Phi$ is not present
in reality: it arises from using DPS approximations in the small $y$
region where they are not valid.  It should hence be removed and replaced
with the appropriate SPS expression, in a manner that avoids double
counting between SPS and DPS.  The analogous double counting problem for
multi-jet production has already been noticed some time
ago\cite{Cacciari:2009dp}.

A short-distance divergence in the $\tvec{y}$ integral also appears for
so-called 2v1 (2 versus 1 parton) diagrams as in figure
\ref{fig:split}(c), where a $1\to 2$ splitting takes place in only one
proton.  This divergence is only logarithmic, and it corresponds to the
overlap of the DPS contribution with the twist-four mechanism shown in
\fref{fig:power}(d).  The importance of the 2v1 mechanism has been
emphasised in Refs.~\refcite{Blok:2011bu,Blok:2013bpa,%
  Ryskin:2011kk,Ryskin:2012qx,Gaunt:2012dd}.

In the following we describe a solution to these problems that was elaborated
in Ref.~\refcite{Diehl:2017kgu}.  Different approaches have been presented in
Refs.~\refcite{Blok:2011bu,Blok:2013bpa,Ryskin:2011kk,Ryskin:2012qx,%
  Manohar:2012pe}, the one of Refs.~\refcite{Blok:2011bu,Blok:2013bpa} being
reviewed in Ref.~\refcite{Blok:book}.  A detailed comparison between them is
given in Ref.~\refcite{Diehl:2017kgu}.  The formalism described here resolves
the double counting problem, retains the concept of double parton
distributions that have a field theoretic definition, and permits the study of
higher-order contributions in a practical way.  The other approaches just
mentioned do not possess all these features.

The first step is to insert the function $\Phi(y \nu)$ into the cross section
formula \eqref{coll-Xsect-master}.  (We insert the square of $\Phi$ for
consistency with the TMD case).  This function regulates the divergences just
discussed by removing the region ${y} \ll 1/\nu$ from what we \emph{define} to
be DPS.  It must satisfy $\Phi(u) \to 0$ for $u\to 0$ and $\Phi(u) \to 1$ for
$u \gg 1$.  Suitable forms are $\Phi(u) = 1-\exp(-u^2/4)$, or a hard cutoff
$\Phi(u) = \Theta(u-b_0)$ with $b_0 = 2e^{-\gamma_E}$ chosen to simplify
analytic expressions.

To avoid double counting between DPS and SPS, and between DPS and the
twist-four mechanism, we introduce subtraction terms in the overall
cross-section:
\begin{align}
  \label{all-Xsect}
\sigma_{\text{tot}} &= \sigma_{\text{DPS}}
   - \sigma_{\text{1v1,pt}} + \sigma_{\text{SPS}}
   - \sigma_{\text{2v1,pt}} + \sigma_{\text{tw4}} \,.
\end{align}
The subtraction terms depend on $\nu$ in such a way that the dependence on
this unphysical parameter cancels on the right-hand side (to the order of
perturbative accuracy of the calculation).  Note that
$\sigma_{\text{SPS}}$ and $\sigma_{\text{tw4}}$ do not depend on $\nu$.
In particular, $\sigma_{\text{SPS}}$ is simply calculated in the usual way
with no modifications.  The 1v1 subtraction $\sigma_{\text{1v1,pt}}$ is
constructed in a simple way by replacing the DPDs in the cross section
formula \eqref{coll-Xsect-master} by the perturbative splitting
approximation \eqref{split-dpd} or its equivalent at higher orders in
$\alpha_s$.  Similarly, $\sigma_{\text{2v1,pt}}$ is obtained by replacing
one of the two DPDs by its splitting approximation and the other one by
its intrinsic short-distance part.  There are some subtleties in choosing
adequate scales $\mu$ in these distributions, especially if the scales
$Q_1, Q_2, Q_h$ of the two DPS subprocesses and of the SPS subprocess are
very different, but we will not dwell on this here.

An appropriate choice for the scale $\nu$ is the minimum of $Q_1$ and
$Q_2$.  With this choice, $\sigma_{\text{DPS}}$ does contain
short-distance contributions for which the DPS approximations are not
valid, but these contributions are removed by the subtraction terms in the
overall cross section.  This is quite similar to choosing factorisation
scales $\mu \sim Q$ in collinear PDFs and DPDs: the parton distributions
then contain virtualities up to the hard scale $Q$, but double counting is
avoided by subtractions in the hard-scattering cross sections.

Let us demonstrate how the prescription works.  At small ${y} \sim 1/Q$,
one has $\sigma_{\text{DPS}} \approx \sigma_{\text{1v1,pt}} +
\sigma_{\text{2v1,pt}}$ by construction (the product of the intrinsic
parts of each DPD gives a power suppressed contribution at small $y$, so
that the absence of a subtraction term $\sigma_{\text{2v2,pt}}$ is no
problem).  One thus has $\sigma_{\text{tot}} \approx \sigma_{\text{SPS}} +
\sigma_{\text{tw4}}$, as is appropriate for the short-distance region.
The dependence on the unphysical cutoff scale $\nu$ cancels between DPS
and the subtraction terms.  At ${y} \gg 1/Q$, the dominant contribution to
$\sigma_{\text{SPS}}$ comes from 1v1 type loops in the region where the
DPS approximations are valid, such that $\sigma_{\text{SPS}} \approx
\sigma_{\text{1v1,pt}}$.  Similarly, we have $\sigma_{\text{tw4}} \approx
\sigma_{\text{2v1,pt}}$.  As a result we obtain $\sigma_{\text{tot}}
\approx \sigma_{\text{DPS}}$, as appropriate.  The construction just
explained is a special case of the general subtraction formalism discussed
in Chap.~10 of Ref.~\refcite{Collins:2011zzd}.

For the scale choice $\nu \sim \min(Q_1, Q_2)$, one can show that the
combination $\sigma_{\text{tw4}} - \sigma_{\text{2v1,pt}}$ in
\eqref{all-Xsect} is subleading compared to $\sigma_{\text{DPD}}$ by a
logarithm $\log(Q/\Lambda)$, where $\Lambda$ is an infrared scale.  This
combination can hence be dropped at leading logarithmic order, which is of
great practical benefit since the computation of the twist-four
contribution is technically quite involved.  For the same scale choice,
one finds that $\sigma_{\text{DPD}}$ includes the appropriate resummation
of large DGLAP logarithms in the 2v1 graphs.

In order to estimate the theoretical uncertainty from missing higher order
terms in this framework, one can vary the parameters $\mu_1$, $\mu_2$ and
$\nu$, similar to how one varies only $\mu$ in the single scattering case.
Note that the variation in $\nu$ of the DPS term alone provides an
order-of-magnitude estimate of SPS graphs containing a double box as in
\fref{fig:power}(b), since it involves the same PDFs, overall coupling
constants and kinematic region (small ${y}$, corresponding to large
transverse momenta and virtualities of internal lines).  An alternative
estimate is provided by the double counting subtraction term
$\sigma_{\text{1v1,pt}}$.  Therefore, a small $\nu$ variation of
$\sigma_{\text{DPS}}$ compared to its central value indicates that
$\sigma_{\text{1v1,pt}}$ and the corresponding loop contribution to
$\sigma_{\text{SPS}}$ are negligible compared to $\sigma_{\text{DPS}}$.
Several scenarios where the $\nu$ variation is reduced in this way were
found in Ref.~\refcite{Diehl:2017kgu}, for instance when the parton pairs
in the relevant DPDs cannot be produced in a single leading-order
splitting (e.g.~$u\bar{d}\,$), or when low $x$ values are probed in the
DPDs.  In such cases, one may justifiably neglect the appropriate
perturbative order of $\sigma_{\text{SPS}}$, together with the 1v1
subtraction term.  Such processes and kinematic regions are the most
promising ones to make useful calculations and measurements of DPS,
especially because there are only few cases for which SPS is computed at
the order containing the double box (essentially only double electroweak
gauge boson production).

Now let us turn to the TMD case, where the pattern of overlaps and
divergences is somewhat different.  In particular, the ultraviolet
divergences in $\sigma_{\text{DPS}}$ associated with 1v1 graphs become
logarithmic rather than a power.  This is related to the fact that DPS and
SPS have the same power behaviour in the small $q_T$ region.  The
inter-parton distance $\tvec{y}_+ = \tvec{y} +
\tfrac{1}{2}(\tvec{z}_1-\tvec{z}_2)$ in the amplitude and its counterpart
$\tvec{y}_- = \tvec{y} - \tfrac{1}{2}(\tvec{z}_1-\tvec{z}_2)$ in the
complex conjugate amplitude (see \fref{fig:split}(a)) are independent
variables now.  When one of these distances is small, the TMDs are
dominated by perturbative $1\to 2$ splitting for the corresponding parton
pair.  The divergent behaviour of $\sigma_{\text{DPS}}$ in the region
where both ${y}_+$ and ${y}_-$ go to zero corresponds to the overlap of
DPS with the SPS double box graph in \fref{fig:power}(b).  The region
where only ${y}_-$ goes to zero corresponds to an overlap with the SPS/DPS
interference graph in \fref{fig:power}(c).

It is clear then that in this case the DPS term must be regulated when
either ${y}_+$ or ${y}_-$ go to zero, as is done in
\eqref{TMD-Xsect-master}.  The SPS/DPS interference terms must also be
regulated for small ${y}_+$ or ${y}_-$, since they overlap with SPS again.
Subtraction terms must be included as appropriate to remove the double
counting, as elaborated in Ref.~\refcite{Diehl:2017kgu}.


\section{Collinear DPDs: evolution}
\label{sec:dglap}

The twist-two operators in the definition of DPDs contain ultraviolet
divergences that require renormalisation.  This leads to the familiar
DGLAP evolution equations of ordinary PDFs, and to corresponding equations
for collinear DPDs.  Taking different scales $\mu_1, \mu_2$ for the
partons with momentum fractions $x_1$ and $x_2$, we have a homogeneous
evolution equation
\begin{align}
  \label{DGLAP-mu1}
\frac{\partial}{\partial \log\mu_1^2}\,
   \prb{R}{F}_{a_1 a_2}(x_1,x_2,\tvec{y}; \mu_1,\mu_2,\zeta)
&= \sum_{b_1} \int_{x_1}^{1-x_2} \frac{dx'_1}{x'_1}\;
   \prb{R}{P}_{a_1 b_1}\Bigl( \frac{x_1}{x'_1};
      \mu_1, \frac{x_1 \zeta}{x_2} \Bigr)\,
\nonumber \\
 &\quad \times
   \prb{R}{F}_{b_1 a_2}(x'_1,x_2^{},\tvec{y}; \mu_1,\mu_2,\zeta)
\end{align}
in $\mu_1$ and its analogue for $\mu_2$.  The two parton pairs with
momentum fractions $x_1$ or $x_2$ evolve separately.  Note that in the
colour singlet sector, both $\prb{1}{F}$ and $\prb{1}{P}$ are $\zeta$
independent, and $\prb{1}{P}$ is the same DGLAP evolution kernel as for
ordinary PDFs.

The interplay of DGLAP evolution with the splitting mechanism described in
\sref{sec:split} has important consequences when one or both momentum
fractions $x_1$, $x_2$ are small\cite{Diehl:2017kgu}.  It can change the
$1/{y}^2$ dependence of the splitting contribution \eqref{split-dpd} into
a much flatter $y$ dependence, which increases the contribution of the
region $y \gg 1/\nu$ in the 1v1 and 2v1 cross sections.  The size of the
effect depends on kinematics and on the parton types involved.

The Fourier integral that converts $F(x_i,\tvec{y})$ into a momentum space
DPD $F(x_i,\tvec{\Delta})$ has a logarithmic divergence at small ${y}$
from the splitting contribution, which requires additional ultraviolet
renormalisation\cite{Diehl:2011yj}.  In the following, we concentrate on
colour singlet distributions ($R=1$) and on equal scales $\mu_1=\mu_2$.
One way to define $F(x_i,\tvec{\Delta})$ is to perform the Fourier
transform in $D= 4-2\epsilon$ dimensions and use ordinary
$\overline{\text{MS}}$ renormalisation for the splitting divergence.  The
resulting evolution equation has an additional inhomogeneous term, which
at LO in $\alpha_s$ reads
\begin{align}
  \label{DGLAP-inhom}
& \frac{\partial}{\partial\log\mu^2}\,
  \prb{1}{F}_{a_1 a_2}(x_1,x_2,\tvec{\Delta}; \mu)
  = \{ \text{homogeneous terms} \}
\nonumber \\
  &\qquad\qquad + \frac{\alpha_s(\mu)}{2\pi}\,
      \prb{1}{P}_{a_0\to a_1 a_2}\biggl( \frac{x_1}{x_1+x_2} \biggr)\,
    \frac{f_{a_0}(x_1+x_2; \mu)}{x_1+x_2} \,,
\end{align}
with the same kernel $\prb{1}{P}_{a_0\to a_1 a_2}$ as in
\eqref{split-dpd}.  The homogeneous terms have the same form as in the
evolution of $F(x_i,\tvec{y})$, with $\tvec{y}$ replaced by
$\tvec{\Delta}$.  This inhomogeneous evolution has been discussed
extensively in the
literature\cite{Kirschner:1979im,Shelest:1982dg,Snigirev:2003cq,
  Gaunt:2009re,Ceccopieri:2010kg}.

Whilst the scheme presented here requires position space DPDs
$F(x_i,\tvec{y})$ for computing cross sections, the momentum space DPDs
have a property that makes their study worthwhile.  At $\tvec{\Delta} =
\tvec{0}$, unpolarised momentum space DPDs satisfy sum
rules\cite{Gaunt:2009re} for the momentum and the flavour quantum numbers
of one of the two partons:
\begin{align}
\sum_{a_2 = q,\bar{q},g}
   \int_0^{1-x_1} \!\! dx_2\; x_2\;
     \prb{1}{F}_{a_1 a_2}(x_1,x_2,\tvec{0})
  &= (1-x_1)\, f_{a_1}(x_1) \,,
\nonumber \\
\int_0^{1-x_1} \!\! dx_2\;
    \Bigl[\ms \prb{1}{F}_{a_1 q}(x_1,x_2,\tvec{0})
            - \prb{1}{F}_{a_1 \bar{q}}(x_1,x_2,\tvec{0}) \ms\Bigr]
  &= N_{a_1 q}\, f_{a_1}(x_1) \,,
\end{align}
where $N_{a_1 q}$ is a combinatorial factor.  The validity of these sum
rules for $\overline{\text{MS}}$ renormalised distributions can be shown
to all orders in perturbation theory\cite{Plossl:2017wjw}.  A relation
between DPDs in momentum and position space can be established by defining
distributions
\begin{align}
\prb{1}{F}_{\Phi}(x_1,x_2,\tvec{\Delta}; \mu,\nu)
  &= \int d^2\tvec{y}\; e^{i \tvec{\Delta} \tvec{y}}\,
    \Phi(y \nu)\; \prb{1}{F}(x_1,x_2,\tvec{y}; \mu) \,,
\end{align}
where the logarithmic singularity at small ${y}$ is removed by the same
regulator function $\Phi$ used in the cross section.  These DPDs and the
$\overline{\text{MS}}$ renormalised ones differ only by the treatment of
the ultraviolet region, so that their difference can be computed in
perturbation theory\cite{Diehl:2017kgu}.  To order $\alpha_s$, one finds
that 
$\prb{1}{F}(x_1,x_2,\tvec{\Delta}; \mu) -
\prb{1}{F}_{\Phi}(x_1,x_2,\tvec{\Delta}; \mu,\mu)$ at
$\tvec{\Delta} = \tvec{0}$ is a calculable function of
$x_1 /(x_1+x_2)$ times the inhomogeneous term in~\eqref{DGLAP-inhom}.

We note that the factorisation formula \eqref{coll-Xsect-master} can be
rewritten in terms of $\int d^2\tvec{\Delta}\;
\prb{R}{F}_{\Phi}(\bar{x}_1,\bar{x}_2,-\tvec{\Delta})\,
\prb{R}{F}_{\Phi}(x_1,x_2,\tvec{\Delta})$.  This has been used to show
that, at leading logarithmic accuracy, the collinear 2v2 and 2v1 cross
sections given in Refs.~\refcite{Blok:2011bu,Blok:2013bpa,%
  Ryskin:2011kk,Ryskin:2012qx,Gaunt:2012dd} are consistent with the
formalism presented here.\cite{Diehl:2017kgu}.


\section{Soft gluons and Sudakov logarithms}
\label{sec:sudakov}

\begin{figure}
\begin{center}
\subfigure[]{\includegraphics[width=0.48\textwidth]{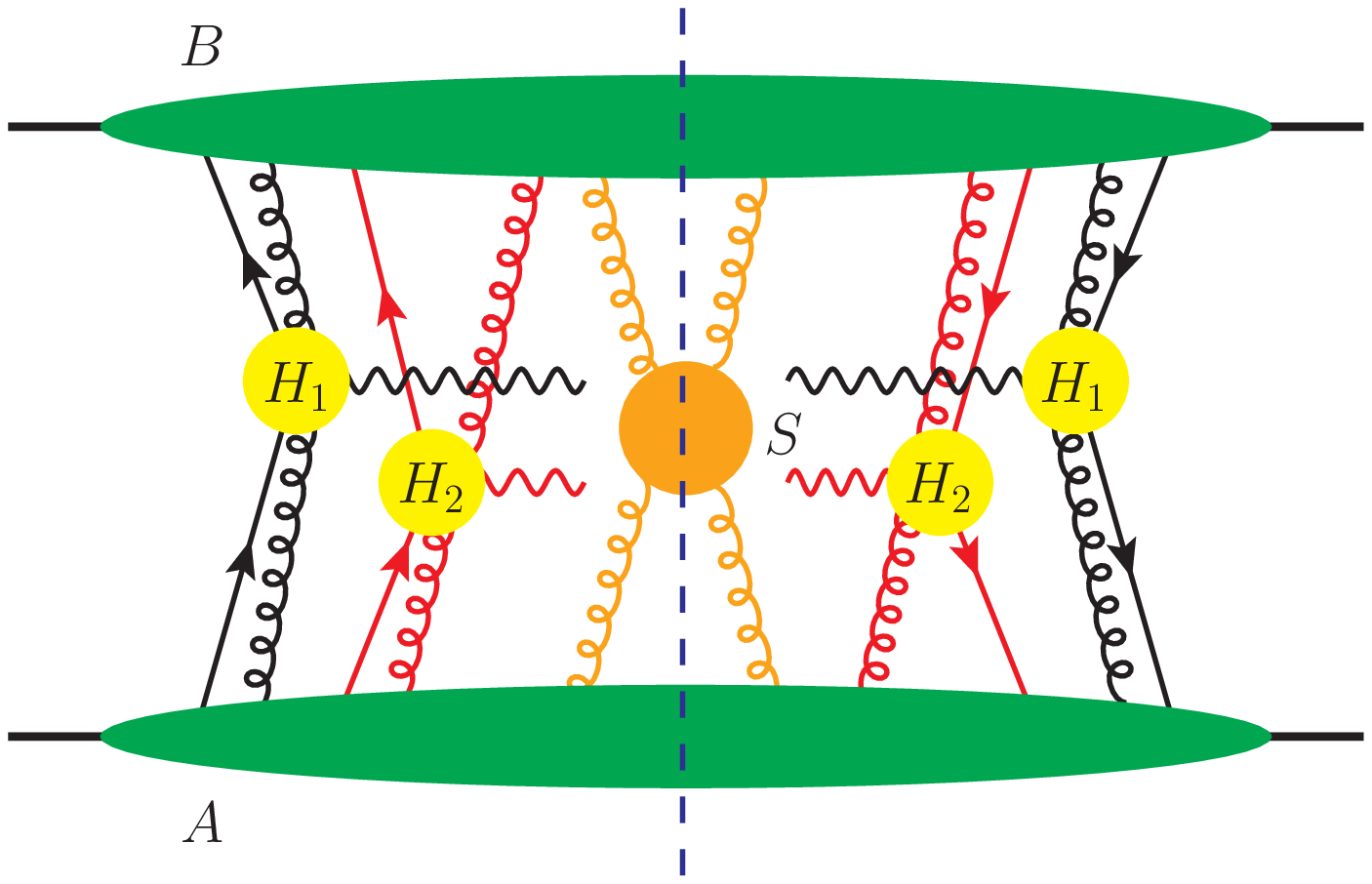}}
\subfigure[]{\includegraphics[width=0.51\textwidth]{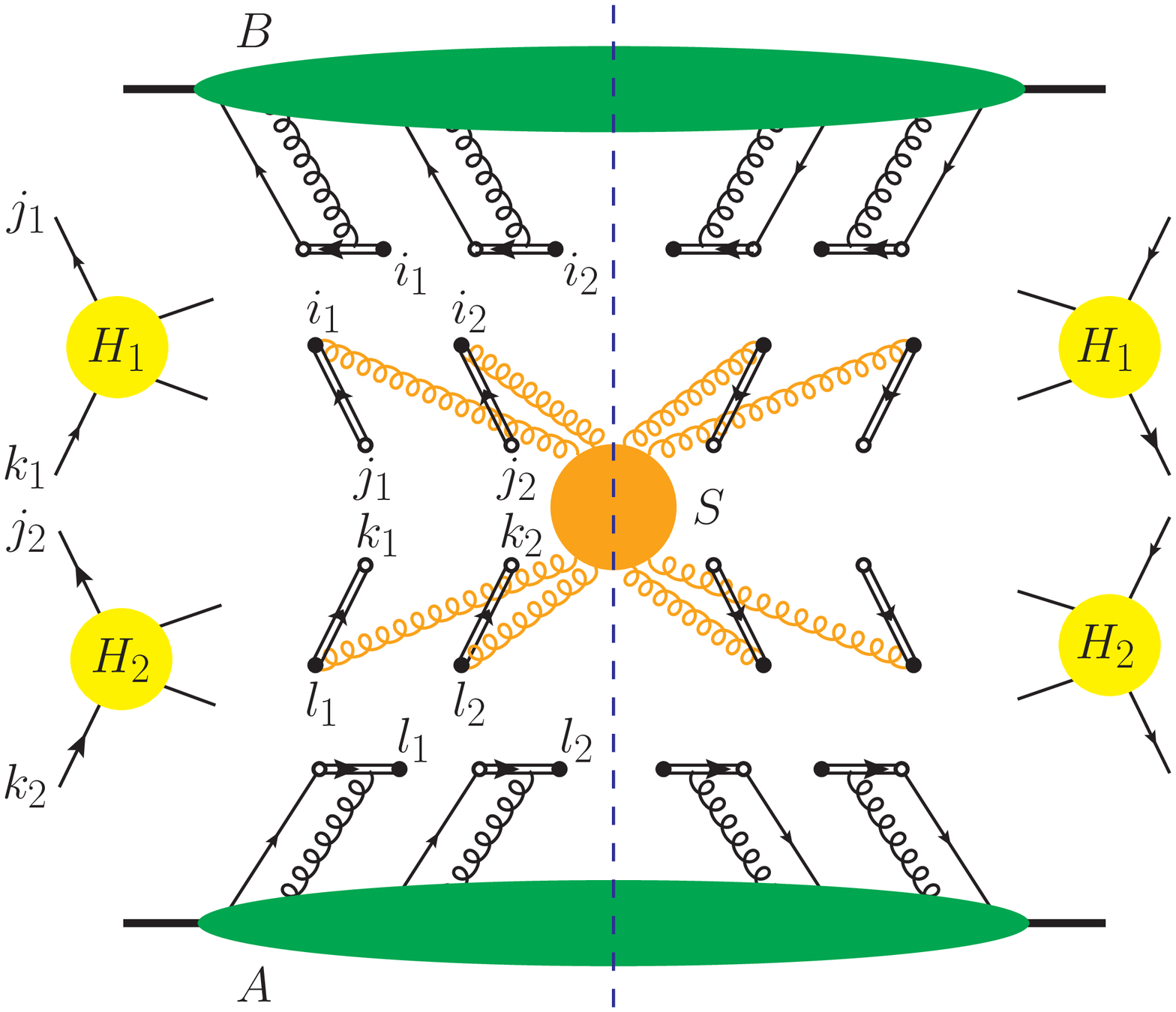}}
\caption{\label{fig:soft-fact} (a) Factorised graph for double Drell-Yan
  production.  (b) Graph for double dijet production after the
  Grammer-Yennie approximation and Ward identities have been applied.
  $i_1, i_2, \cdots, l_2$ are colour indices.}
\end{center}
\end{figure}

The proof of the DPS factorisation formulae \eqref{coll-Xsect-master} and
\eqref{TMD-Xsect-master} proceeds in close analogy to the case of SPS.
Here we only sketch the steps that lead to the construction of DPDs and
their evolution equations in rapidity, referring to
Refs.~\refcite{Diehl:2015bca} and \refcite{Buffing:2017mqm} for details.
One starts by showing that graphs contributing to the cross section at
leading power in $\Lambda/Q$ factorise into hard, collinear and soft
subgraphs, as depicted for the double Drell-Yan process in
\fref{fig:soft-fact}(a).  In the hard-scattering subgraphs $H_1$ and $H_2$
all internal lines are far off shell, the subgraphs $A$ and $B$ involve
only momenta collinear to one of the incoming protons, and the subgraph
$S$ describes the exchange of soft gluons between the right-moving partons
in $A$ and the left-moving ones in $B$.

To obtain a useful factorisation formula, soft gluons must be decoupled
from $A$ and $B$.  To achieve this, one performs a Grammer-Yennie
approximation.  For one gluon with momentum $\ell$ flowing from $S$ into
$A$, this reads
\begin{align}
  \label{gram-yen}
S_\mu(\ell)\, A^\mu(\ell)
 & \approx S^-(\ell)\, \frac{v_R^+}{\ell^-_{\phantom{R}} v_R^+ + i\epsilon}\, 
    \ell^-\! A^+(\tilde{\ell})
 \approx S_\mu(\ell)\,
  \frac{v_R^\mu}{\ell v_R^{} + i\epsilon}\, \tilde{\ell}_\nu
  A^\nu(\tilde{\ell}) \,,
\end{align}
where $v_R^{} = (v_R^+,v_R^-,\tvec{0})$ specifies a direction with large
positive rapidity $Y_R = \half \log|v_R^+/v_R^-|$.  In the first step, we
have retained only the largest component of $A^\mu$ and replaced $\ell =
(\ell^+,\ell^-,\tvec{\ell})$ with $\tilde{\ell} = (0,\ell^-,\tvec{\ell})$
in that factor, which involves plus momenta much larger than $\ell^+$.
The second step is more delicate and will be discussed in
\sref{sec:glauber}.  One can now apply a Ward identity to
$\tilde{\ell}_\nu  A^\nu(\tilde{\ell})$, which removes the gluon
attachment from $A$.  The soft gluon emerging from $S$ then couples to a
Wilson line along the direction $v_R$.  Naively, one would take $v_R$
lightlike, i.e.\ set $v_R^-=0$, but this would lead to so-called rapidity
divergences of loop integrals inside $S$.  Instead of a finite rapidity
$Y_R$, one may use other methods to regulate these divergences, two of
which have been applied to DPS in Refs.~\refcite{Manohar:2012jr} and
\refcite{Vladimirov:2016qkd}.

In full analogy, one can replace soft gluons coupling to $B$ by gluons
coupling to a Wilson line along a direction $v_L$ with large negative
rapidity.  The soft factor $S$ can now be written as the vacuum
expectation value of Wilson line operators and is thus decoupled from $A$
and $B$.

A similar argument is used for longitudinally polarised collinear gluons
exchanged between $A$ and $H_1$ or $H_2$, removing them from the hard
scatters and coupling them to Wilson lines along $v_L$.  These Wilson
lines appear in the proton matrix element
$\langle p \ms| \mathcal{O}_{a_1} \mathcal{O}_{a_2}|\ms p\rangle$
associated with the factor $A$.  The same steps are followed for gluons
exchanged between $B$ and $H_1$ or $H_2$, resulting in Wilson lines along
$v_R$ in the matrix element associated with~$B$.

The factors in the cross section formula are tied together by colour
indices in a way that is shown in \fref{fig:soft-fact}(b) for double dijet
production.  We first discuss the production of colour singlet particles
in $H_1$ and $H_2$.  In this case, the soft factor is contracted with
$\delta^{j_1 k_1} \delta^{j_2 k_2}$ and analogous factors for the index
pairs on the r.h.s.\ of the final-state cut.

It is useful to project the collinear and soft factors on colour
representations $R$, as was explained for DPDs in \sref{sec:cross-sect}.
The collinear factors then become vectors $\prb{R}{\bs A}$, $\prb{R}{B}$
in the space of colour representations, and the soft factor for colour
singlet production becomes a matrix $\prl{RR'}{S}$.  In the cross section
we have the combination $\sum_{R R'} \prb{R}{B}\; \prl{RR'}{S}\;
\prl{R'\!}{A}$.

The soft matrix in the cross section depends on the rapidity difference $Y
= Y_R-Y_L$ of the Wilson lines along $v_R$ and $v_L$.  This dependence is
given by a Collins-Soper equation
\begin{align}
  \label{soft-cs}
\frac{\partial}{\partial Y}\, \prl{RR''}{S}(Y) &=
  \sum_{R'} \prl{RR'}{\widehat{K}}\,\, \prl{R'\! R''}{S}(Y)
\end{align}
with a rapidity independent matrix ${\widehat{K}}$.  Introducing an
intermediate rapidity $Y_C$ between $Y_R$ and $Y_L$, one can write $S(Y)$
as the product of two matrices, one depending on $Y_R-Y_C$ and the other
on $Y_C-Y_L$.  Combining one of them with $A$ and the other with $B$, one
obtains DPDs ${F}_{A}(Y_C)$ and ${F}_{B}(Y_C)$.  In the DPDs, one can take
the limits $Y_L\to -\infty$ and $Y_R\to \infty$ without encountering
rapidity divergences.  The proton matrix element in~\eqref{tmd-dpds} then
contains lightlike Wilson lines.
The final cross section formula involves the sum $\sum_{R} \prb{R}{F}_B\,
\prb{R}{F}_A$ with colour dependent DPDs, as anticipated in
\sref{sec:cross-sect}.  Note that $A$, $B$ and $S$ are nonperturbative
quantities since they involve low virtualities.  Eliminating $S$ by
defining distributions $F_A$ and $F_B$ thus represents a significant
simplification.

The construction sketched so far is common to TMD and collinear
factorisation.  However, ultraviolet renormalisation works differently in
the two cases, which we now discuss in turn.


\paragraph{TMD factorisation.}  
It is useful to express the rapidity dependence of the DPDs in terms of
boost invariant quantities, $\zeta = 2 x_1 x_2 (p^+)^2\, e^{-2 Y_C}$ for
$F_A$ and an analogue $\bar{\zeta}$ for $F_B$.  We concentrate on $F_A$
from now on and omit the subscript $A$.  Restoring the dependence on all
other variables, we have a Collins-Soper equation
\begin{align}
  \label{tmd-cs}
  \frac{\partial}{\partial\log\zeta}\ms
  \prb{R}{F}(x_i,\tvec{z}_i,\tvec{y};\mu_i,\zeta) &= \frac{1}{2} \sum_{R'}
  \prl{RR'}{K}(\tvec{z}_i,\tvec{y};\mu_i)\,
  \prl{R'}{F}(x_i,\tvec{z}_i,\tvec{y};\mu_i,\zeta)
\end{align}
with a matrix kernel $K$ related to $\widehat{K}$ in \eqref{soft-cs}.  Its
$\mu_1$ dependence is given by
\begin{align}
  \label{cs-rge}
\frac{\partial}{\partial\log\mu_1}\,
  \prl{RR'}{K}(\tvec{z}_i,\tvec{y};\mu_i)
   &= {}- \delta_{RR'}\, \gamma_K(\mu_1) \,,
\end{align}
whilst for the DPD we have
\begin{align}
  \label{tmd-rge}
\frac{\partial}{\partial\log\mu_1}\,
  \prb{R}{F}(x_i,\tvec{z}_i,\tvec{y};\mu_i,\zeta)
&= \gamma_{F}(\mu_1, x_1 \zeta/x_2)\,
    \prb{R}{F}(x_i,\tvec{z}_i,\tvec{y};\mu_i,\zeta)
\end{align}
with
\begin{align}
\gamma_{F}(\mu, \zeta) &= \gamma_F(\mu,\mu^2) - \frac{1}{2}\ms
      \gamma_K(\mu)\, \log\frac{\zeta}{\mu^2} \,.
\end{align}
Here $\gamma_K(\mu)$ and $\gamma_F(\mu,\mu^2)$ depend
on $\mu$ via $\alpha_s(\mu)$.  Analogues of \eqref{cs-rge} and
\eqref{tmd-rge} hold for the $\mu_2$ dependence.  Note that the kernel $K$
and the anomalous dimensions $\gamma_K$
and $\gamma_F$ differ for quarks and for gluons, but are independent of
parton polarisation and quark flavour.

The system of evolution equations can be solved analytically (provided the
matrix $K$ can be diagonalised analytically).  The solution exponentiates
Sudakov double logarithms, controlled by $\gamma_K$, and single logarithms
going with $\gamma_F$ and $K$.  Except for the scaling of $\zeta$ by
$x_1/x_2$ or $x_2/x_1$, the double logarithms have the same form as for
single TMDs.  When $F_A$ is multiplied with $F_B$ in the cross section,
logarithms of $\zeta$ and $\bar{\zeta}$ turn into logarithms of the
invariant masses $Q_1$ and $Q_2$ in the two hard-scattering subprocesses.


\paragraph{Collinear factorisation.}
In collinear factorisation, the soft factor simplifies considerably.
Using colour algebra, one can show that the general soft factor shown in
\fref{fig:soft-fact}(b) is the same as the one for colour singlet
production, which is contracted with $\delta^{j_1 k_1} \delta^{j_2 k_2}$
etc.  Moreover one finds that the soft matrix $\prl{RR'}{S}$ is diagonal.
The colour matrix algebra in the construction of DPDs thus becomes
trivial.  The product of factors in the cross section can then be written
as
\begin{align}
\sum_R \prb{R}{H}_1\ms \prb{R}{H}_2\ms \prb{R}{F}_B\ms \prb{R}{F}_A \,.
\end{align}
If a colour singlet state is produced, then all colour projections
$\prb{R}{H}$ are equal, otherwise they differ.  Multiplying $\prb{R}{H}$
with a flux factor, one obtains the subprocess cross sections
$\prn{R}{\hat{\sigma}}$ in the collinear factorisation formula
\eqref{coll-Xsect-master}.  Collinear DPDs depend on $\zeta$ as
\begin{align}
  \label{coll-cs}
  \frac{\partial}{\partial\log\zeta}\,
  \prb{R}{F}(x_i,\tvec{y};\mu_i,\zeta)
  &= \frac{1}{2}\, \prb{R}{J}(\tvec{y};\mu_i)\,
     \prb{R}{F}(x_i,\tvec{y};\mu_i,\zeta)
\end{align}
with
\begin{align}
\frac{\partial}{\partial\log\mu_1}\, \prb{R}{J}(\tvec{y};\mu_i)
  &= {}- \prn{R}{\gamma}_J(\mu_1)
\end{align}
and an analogous equation for the $\mu_2$ dependence.  The DGLAP kernels
in the evolution equation \eqref{DGLAP-mu1} depend on $\zeta$ via
\begin{align}
\prb{R}{P}_{a b}(x;\mu,\zeta)
  &= \prb{R}{P}_{a b}(x;\mu,\mu^2)
    - \frac{1}{4}\ms \delta_{a b}\ms \delta(1-x)\,
    \prn{R}{\gamma}_J(\mu)\, \log\frac{\zeta}{\mu^2} \,.
\end{align}
The $\zeta$ dependence of the DPDs can be given in analytical form.  It
contains exponentiated double logarithms controlled by $\gamma_J$ and
single logarithms going with the kernel $J$.  In the colour singlet
sector, the soft factor is trivial, $\prb{11}{S} = 1$, and correspondingly
one has $\prb{1}{J} = \prn{1}{\gamma}_J = 0$.  In physical terms, the
effects of soft gluon exchange cancel in this case.  As a result, colour
non-singlet DPDs are suppressed by Sudakov logarithms, whereas colour
singlet DPDs are not\cite{Mekhfi:1988kj,Manohar:2012jr}.


\section{Glauber gluons and factorisation}
\label{sec:glauber}

A crucial step in showing that soft gluon exchange between the subgraphs
$A$ and $B$ in \fref{fig:soft-fact}(a) can be subsumed into the vacuum
expectation value of Wilson lines is to establish the absence of
contributions from the so-called Glauber region.  In the following, we
restrict ourselves to colour singlet production, since it is the only
context in which this issue has been studied.  The arguments apply both to
collinear and TMD factorisation.

For gluons leaving the soft subgraph $S$ in \fref{fig:soft-fact}(a),
there are in fact two distinct momentum regions that contribute to the
cross section at leading power.  The first one can be called the ``central
soft'' region, where all components of the momentum $\ell$ have comparable
size, $|\ell^+| \sim |\ell^-| \sim |\tvec{\ell}|$.  In this region, the
second step of the Grammer-Yennie approximation \eqref{gram-yen} is valid:
we have $\ell^- \! A^+ \approx \tilde{\ell}_\nu A^\nu$ because
$\tvec{\ell}\ms \tvec{A}$ is power suppressed compared to $\ell^- \!
A^+$.
The second one is the Glauber region, which is characterised by
$|\ell^+\ell^-| \ll \tvec{\ell}^2$.  Gluons in this region mediate
small-angle scattering of a right-moving parton on a left-moving one.  In
the Glauber region, we can have $|\tvec{\ell}\ms \tvec{A}| \sim |\ell^-
A^+|$, so that the Grammer-Yennie approximation fails.  This presents a
serious obstacle to factorisation.

Of course, the soft momentum $\ell$ in a graph is not held fixed but
integrated over.  For many types of soft gluon attachment, the integration
over $\ell^+$ or $\ell^-$ (or both) can be deformed away from the real
axis into the complex plane in such a way that one has
$|\ell^+\ell^-| \sim \tvec{\ell}^2$ on the deformed integration contour
and thus avoids the Glauber region.  This is only possible when the poles
in $\ell^+$ or $\ell^-$ of the propagators depending on $\ell$ do not
obstruct the deformation.  In such cases, the contribution from $\ell$ in
the Glauber region can be validly subsumed into the contribution from a
collinear or a central soft region, where Grammer-Yennie approximations
can be applied to achieve factorisation.

\begin{figure}
\begin{center}
\includegraphics[width=0.57\textwidth]{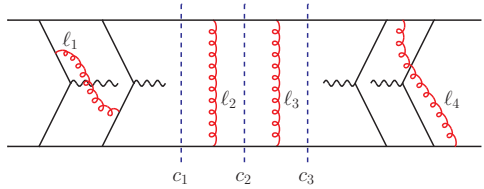}
\caption{\label{fig:glauber} Graph for double Drell-Yan production with
  several soft gluons exchanged between left and right moving fast
  partons.  The three possible final state cuts of the graph are denoted
  by $c_1, c_2$ and $c_3$.}
\end{center}
\end{figure}

Examples of soft attachments for which we may deform the momentum out of
the Glauber region are given by the gluons with momenta $\ell_1$ and
$\ell_4$ in \fref{fig:glauber}.  In fact, for all of the ``novel'' types
of soft attachment that appear only when we consider DPS rather than SPS,
such a deformation in possible.  Note that the sign of $i\epsilon$ in the
denominators of \eqref{gram-yen} is chosen precisely such that it does not
obstruct these contour deformations.  As a consequence, the Wilson lines
in the construction of soft factors and DPDs are past pointing (in
light-cone coordinates), as in \eqref{WL-def}.

The one type of soft attachment for which the propagator poles obstruct a
deformation out of the Glauber region is exemplified by the gluons with
momenta $\ell_2$ and $\ell_3$ in \fref{fig:glauber}.  This is an attachment
between a right-moving and a left-moving spectator parton after the two hard
scatters (where ``after'' refers to the topology of graphs and not to the time
coordinate in some reference frame).  Of course, such exchanges occur already
in SPS.  As shown for instance in Ref.~\refcite{Gaunt:2014ska}, the
contribution from the Glauber region cancels to leading power when one sums
over all final state cuts of a given graph.  This requires that the cross
section is differential only in properties of the hard scattering products,
but fully inclusive over the remaining particles.  The same argument applies
to DPS.  The principle ensuring this cancellation is unitarity: spectator
scattering does affect details of the final state, but its net effect is zero
if the observable is not sensitive to the spectator momenta.  The DPS cross
sections \eqref{coll-Xsect-master} and \eqref{TMD-Xsect-master} satisfy this
requirement.  On the other hand, one can show that Glauber gluon exchange does
break factorisation for observables depending on the momenta of the spectator
partons (or better, of the ``beam jets'' into which these partons hadronise).

The argument just sketched works for simple graphs, essentially at the level
of single soft gluon exchange.  To demonstrate Glauber cancellation at all
orders, a more powerful technique is needed, based on the light-front ordered
version of QCD perturbation theory (LCPT).  This argument was given for DPS in
Ref.~\refcite{Diehl:2015bca}, generalising the treatment of the SPS case in
Ref.~\refcite{Collins:2011zzd}.  In the LCPT picture, one sees again that from
the point of view of the Glauber gluons, single and double hard scattering
look rather similar, and that the troublesome ``final state'' poles
obstructing the deformation out of the Glauber region cancel after the sum
over of final state cuts.  Again, a unitarity argument is used to achieve this
cancellation.


\section{Perturbative transverse momenta}
\label{sec:high-qt}

TMD factorisation is applicable when the typical size $q_T$ of the transverse
momenta $\tvec{q}_1, \tvec{q}_2$ is much smaller than the hard scale $Q$.
This includes the multi-scale regime $\Lambda \ll q_T \ll Q$.  In this
situation, the small parameter in the general power counting of
\sref{sec:power} becomes $q_T/Q$ rather than $\Lambda/Q$.

The Fourier exponent
$e^{- i (\tvec{q}{}_1 \tvec{z}_1 + \tvec{q}{}_2 \tvec{z}_2)}$ in the TMD
cross section \eqref{TMD-Xsect-master} limits the distances $z_1$ and
$z_2$ to typical size $1/q_T$.  Double TMDs can then be computed in terms
of perturbative subprocesses at scale $q_T$ and of collinear matrix
elements expressing the physics at scale $\Lambda$, which significantly
increases the predictive power of theory.  The distance $y$ is not
restricted in this way, and there are in fact two different regimes for
DPS.

\begin{figure}[t]
\begin{center}
\subfigure[]{\includegraphics[height=0.14\textwidth]{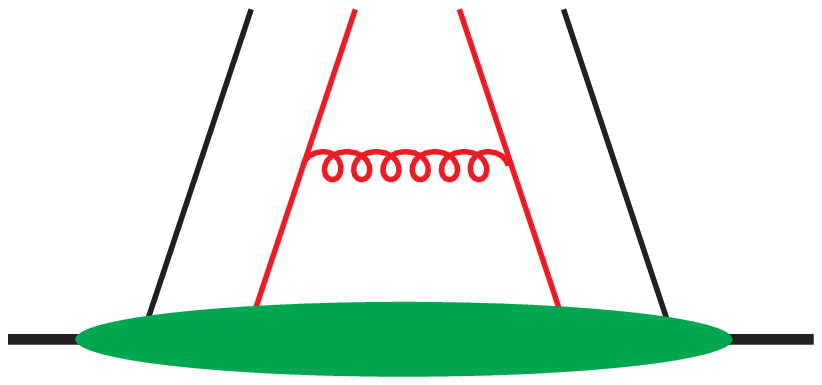}}
\hspace{2em}
\subfigure[]{\includegraphics[height=0.14\textwidth]{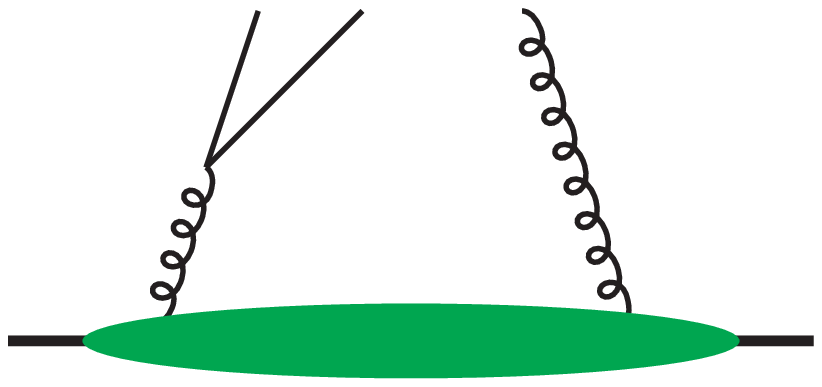}}
\caption{\label{fig:dpd-short} (a) Ladder graph for a DPD.  (b)
  Splitting contribution for a twist-three TMD.}
\end{center}
\end{figure}

If $y \sim 1/\Lambda$ is of hadronic size, then the mechanism generating
perturbative transverse momenta is the emission of partons, described by
ladder graphs as in \fref{fig:dpd-short}(a).  For the DPDs we then have
\begin{align}
  \label{tmd-match}
& \prb{R}{F}_{a_1 a_2}(x_i,\tvec{z}_i,\tvec{y};\mu_i,\zeta)
 = \sum_{b_1 b_2}
  \prn{R}{C}_{a_1 b_1}(x_1',\tvec{z}_1^{};\mu_1^{}, x_1^{} \zeta/x_2^{})
\nonumber \\
 &\qquad \underset{x_1}{\otimes}
  \prn{R}{C}_{a_2 b_2}(x_2',\tvec{z}_2^{};\mu_2^{}, x_2^{} \zeta/x_1^{})
  \underset{x_2}{\otimes}
  \prb{R}{F}_{b_1 b_2}(x_i',\tvec{y};\mu_i^{},\zeta)
\end{align}
with convolution products $\otimes$ as in \eqref{DGLAP-mu1}.  This is the same
mechanism as in SPS (a prominent example is Drell-Yan production with
$\Lambda \ll q_T \ll Q$), and the short-distance coefficients $\prn{1}{C}$ are
the same as the ones for single TMDs.  The relation \eqref{tmd-match} can be
understood in terms of an operator product expansion, with the operators in
\eqref{quark-ops} being expanded for small $z_1, z_2$ while keeping $y$ large.

The second regime for DPS is when $y$ is of the same size as $z_i \sim 1/q_T$.
The operator product expansion for the double TMD then involves three types of
terms, already presented in \sref{sec:split}.  The four partons at small
relative distances can originate from a collinear PDF via splitting, as in
\fref{fig:split}(a), from a collinear twist-four distribution without any
splitting, or from a collinear twist-three distribution with parton splitting
only in the amplitude or its conjugate.  As already noted, the latter case
requires chiral-odd distributions and lacks small $x$ enhancement.

In the DPS cross section, we have 1v1, 2v1 and 2v2 contributions from the
four combinations of splitting and intrinsic contributions to the two
DPDs.  It is important to note that these have different power behaviour
in $q_T$, namely
\begin{align}
  \label{tmd-powers}
\frac{Q^4\ms d\sigma}{d^2\tvec{q}_1\ms d^2\tvec{q}_2}
 & \underset{y q_T\sim 1}{\sim} \begin{cases}
    \alpha_s^2/q_T^2 & \text{for 1v1} \\
    \alpha_s\ms \Lambda^2/q_T^4 & \text{for 2v1} \\
    \Lambda^4/q_T^6 & \text{for 2v2}
\end{cases}
& \quad
\frac{Q^4\ms d\sigma}{d^2\tvec{q}_1\ms d^2\tvec{q}_2}
 & \underset{y\Lambda\sim 1}{\sim} \Lambda^2/q_T^4 \,,
\end{align}
where we have also specified the behaviour of the contribution from
$y\sim 1/\Lambda$.  Although this contribution, as well as the 2v1 part at
$y\sim 1/q_T$ are suppressed by $\Lambda^2/q_T^2$ compared with 1v1, it makes
sense to keep them since they have a stronger small $x$ enhancement and
involve fewer powers of $\alpha_s$.  Explicit expressions for the different
terms, including the Sudakov factors resulting from $\zeta$ evolution, are
given in Ref.~\refcite{Buffing:2017mqm}. We note that \eqref{tmd-powers}
holds if $|\tvec{q}{}_1 + \tvec{q}{}_2| \sim |\tvec{q}{}_1| \sim
|\tvec{q}{}_2|$ are all of order $q_T$.  Other regimes have been discussed
in Refs.~\refcite{Diehl:2011yj,Blok:2011bu,Blok:2013bpa}.

To obtain the physical cross section, one must combine DPS with SPS and
the SPS/DPS interference, as discussed in \sref{sec:split}.  The TMDs in
these contributions can be expressed in terms of collinear matrix elements
as well.  For SPS, they are just the ordinary PDFs.  For the interference
term, one has contributions with collinear twist-three distributions
(lacking small $x$ enhancement) and contributions with a PDF and a
short-distance splitting only on one side of the final state cut, as shown
in \fref{fig:dpd-short}(b).  Overall, one thus finds that -- if collinear
twist-three distributions are neglected -- the only parton distributions
needed for TMD factorisation in the regime $\Lambda \ll q_T \ll Q$ are
collinear DPDs and ordinary PDFs.


\section{Status of factorisation}
\label{sec:status}

Significant progress has been made towards establishing factorisation formulae
for DPS processes at the same level of rigour as for SPS.  In fact, many of
the results we have sketched can even be extended to the case of three or more
hard scatterings in a rather straightforward manner.  However, a description
of the colour structure becomes rather cumbersome in this case, as does the
discussion of perturbative splitting and double counting with other
mechanisms.  To conclude this overview, we list what in our opinion are major
remaining open issues in DPS factorisation.

No all-order proof is available for the nonabelian Ward identities required
for decoupling soft gluons from the collinear factors (see
\sref{sec:sudakov}).  Examples at lowest order have been given in
Ref.~\refcite{Diehl:2011yj}.  It may be possible to adapt the proof of the
corresponding Ward identities in single Drell-Yan
production\cite{Collins:1988ig} to DPS, but this has not been worked out.

A crucial ingredient for constructing DPDs is the evolution
equation~\eqref{soft-cs} of the soft matrix $S$, for which no general proof
has been given yet.  For small distances $y$ and $z_i$, one can calculate $S$
in perturbation theory and easily finds that \eqref{soft-cs} is valid at one
loop\cite{Diehl:2011yj}.  Its validity at two loops is corroborated by the
calculation in Ref.~\refcite{Vladimirov:2016qkd} (which uses a different
regulator for rapidity divergences).  An all-order proof has recently been put
forward in Ref.~\refcite{Vladimirov:2017ksc}, but it is currently not clear
whether it applies to the rapidity regulator employed in the present work.
We also note that the construction sketched below~\eqref{soft-cs} requires $S$
to be positive semidefinite.  There is no general proof for this, but it can
be motivated by perturbative arguments\cite{Buffing:2017mqm}.

A technical problem in the construction of soft factors are gluons that
couple only to Wilson lines along one direction.  Such so-called Wilson
line self-interactions are divergent for Wilson lines of infinite length.
It is easy to see that they cancel in the cross section by construction,
but one must also show that they cancel in the individual parton
distributions in the factorisation formula.  Some discussion for SPS is
given in Chap.~13.7 of Ref.~\refcite{Collins:2011zzd}, but it would be
desirable to have a more explicit solution to this problem, before
applying it to the case of DPS.

Finally, the cancellation of Glauber gluon exchange has only been shown
for DPS processes producing colourless particles\cite{Diehl:2015bca}.  An
extension of this argument to the production of coloured particles,
relevant e.g.\ for jet production, has not even been worked out for SPS,
as far as we know.  Such an extension may be possible for collinear
factorisation, whereas for TMD factorisation there are strong arguments
that this cannot even be done for SPS\cite{Rogers:2010dm}.

Many of the subtleties in DPS factorisation, such as the presence of
parton correlations and the perturbative splitting mechanism, are by now
quite well understood on the theory side.  Their phenomenological
importance, however, remains to be quantified for many interesting cases.
This opens a wide field of studies for the future.


\section*{Acknowledgements}

J.G.\ acknowledges financial support from the European Community under the
FP7 Ideas program QWORK (contract 320389).  The figures in this
contribution were produced with JaxoDraw\cite{Binosi:2008ig}.


\bibliographystyle{JHEP}

\bibliography{dps-theory}


\end{document}